\documentclass[%
 reprint,
 amsmath,amssymb,
 aps,
]{revtex4-2}
\usepackage{mathrsfs}
\usepackage{color}
\usepackage{ulem}
\usepackage{graphicx}
\usepackage{dcolumn}
\usepackage{bm}

\newcommand{\tk}[1]{\textcolor{black}{#1}}
\newcommand{\Erase}[1]{\if0{#1}\fi}

\newcommand{\editage}[1]{\textcolor{black}{#1}}

\begin{document}

\preprint{APS/123-QED}

\title{Transport properties \Erase{in}\editage{of} organic Dirac electron system $\alpha$-(BEDT-TSeF)$_2$I$_3$}

\author{D. Ohki$^1$}
\email{dohki@s.phys.nagoya-u.ac.jp}
\author{K. Yoshimi$^2$}%
\author{A. Kobayashi$^1$}
 \affiliation{$^1$Department of Physics, Nagoya University, Furo-cho, Chikusa-ku, Nagoya, 464-8602 Japan \\
  $^2$Institute for Solid State Physics, University of Tokyo, Chiba 277-8581, Japan \\
}%




\date{\today}

\newcommand{\ohki}[1]{\textcolor{black}{#1}}

\begin{abstract}
Motivated by the insulating behavior of \tk{$\alpha$-(BEDT-TSeF)$_2$I$_3$ at low temperatures ($T$s)} , we first performed first-principles calculations based on the crystal structural data at 30 K under ambient pressure and \tk{constructed a two-dimensional effective model using \Erase{M}\editage{m}aximally localized Wannier functions.} 
As \Erase{candidates for the}\editage{possible} cause\editage{s} of the insulating behavior, we studied the effects of the on-site Coulomb interaction $U$ and spin-orbit interaction (SOI) by \tk{investigating} the electronic state and the transport coefficient using the Hartree approximation and the $T$-matrix approximation. 
The calculations at a finite $T$ demonstrated that spin-ordered massive Dirac electron (SMD) appeared owing to the on-site Coulomb interaction.
\tk{
We had an interest in the anomalous competitive effect with $U$ and SOI when the SMD phase is present in $\alpha$-(BETS)$_2$I$_3$ and investigated these contribution to the electronic state and conductivity.
}
SMD is not a conventional spin order, but exhibits the spin-valley Hall effect.
Direct current resistivity \tk{in the presence of a spin order gap} divergently increased and exhibited negative magnetoresistance \tk{in the low $T$ region with decreasing $T$}.
The charge density hardly changed below and above the $T$ at which this insulating behavior appeared. 
However, when considering the SOI alone, the \tk{state changed to a topological insulator phase}, and the electrical resistivity \tk{is} saturated by edge conduction at quite low $T$.
When considering both the SMD and the SOI, the spin order gap was suppressed by the SOI, and gaps with different sizes opened in the left and right Dirac cones. 
This phase transition leads to distinct changes in microwave conductivity, such as a discontinuous jump and a peak structure. 
\end{abstract}

\maketitle


\section{\label{sec:level1}Introduction}

Quasiparticles that have properties similar to \editage{those of} relativistic particles in solids have been found in various materials such as graphene \cite{Wallace, Novoselov}, bismuth \cite{Wolff, FukuyamaKubo}, and \tk{several} organic conductors \cite{Kajita1992, Tajima2000, Kobayashi2004, Katayama2006, Kobayashi2007, Goerbig2008, Kajita2014, Tajima2006}. 
They are called Dirac electrons in solids and exhibit exotic physical properties such as quantum transport \cite{Shon}. 
\Erase{In}\editage{For} Dirac electrons in organic conductors such as $\alpha$-(BEDT-TTF)$_2$I$_3$ and $\alpha$-(BEDT-TSeF)$_2$I$_3$ ($\alpha$-(BETS)$_2$I$_3$), which are the main focus in this study, the Coulomb interaction is relatively large owing to the narrow band width. 
The relationship between the Dirac electron and the electron correlation effect has been discussed. 

In $\alpha$-(BEDT-TTF)$_2$I$_3$, it is suggested that phase transition between the Dirac electron phase and the charge-ordered insulator phase is induced by the nearest-neighbor Coulomb interaction \cite{Seo2000, Takahashi, Kakiuchi}, and anomalous behaviors associated with the electron correlation effect such as pressure dependence of the spin gap \cite{TanakaOgata, Ishikawa} and transport phenomena at low temperatures ($T$s) \cite{Beyer, Liu, Ohki} have been observed. 
It has also been shown that a long-range component of the Coulomb interaction induces reshaping of the Dirac cone \cite{HirataNature, Matsuno2018}, and it enhances spin-triplet excitonic fluctuations in the massless Dirac Electron phase under high pressure and in-plane magnetic field \cite{HirataScience}. 

 $\alpha$-(BETS)$_2$I$_3$ is a related substance of $\alpha$-(BEDT-TTF)$_2$I$_3$. 
In the composition of the BETS molecule, the \tk{Sulfur (S)} atom in the BEDT-TTF molecule is replaced with a \tk{Selenium (Se)} atom, and its relationship with the high-pressure phase of $\alpha$-(BEDT-TTF)$_2$I$_3$ has been discussed. 
Direct current (DC) electrical resistivity measurements showed that properties of Dirac electron appear at $T > 50$ K~\cite{Inokuchi}. 
On the other hand, at $T < 50$ K, the DC resistivity increases divergently. 
Nuclear magnetic resonance (NMR) measurements indicated that \tk{an energy} gap \tk{$\sim$ 300K} is opened at low $T$~\cite{Hiraki}. 
However, unlike in the $\alpha$-(BEDT-TTF)$_2$I$_3$, the inversion symmetry is not broken \tk{and charge density at each site hardly changes in $30$ K $< T < 80$ K}, which has been revealed recently by the synchrotron X-ray diffraction experiment~\cite{KitouSawaTsumuraya}. 
Thus, \tk{the insulation mechanism of $\alpha$-(BETS)$_2$I$_3$ is not related to the charge order, and }the electronic state at low $T$ has not been clarified.

Under hydrostatic pressure, the energy band with electron and hole pockets is obtained by band calculations using the extended H\"{u}ckel method or first-principles calculation~\cite{Kondo, Alemany}. 
A mean-field calculation using the extended Hubbard model based on the extended H\"{u}ckel method suggests that the insulating state at low $T$ is a band insulator due to merging of the Dirac cones \cite{MorinariSuzumura}. 
However, high-accuracy X-ray diffraction data at 30 K under ambient pressure have recently been obtained, and \Erase{as a result of}\editage{using} first-principles calculation, it has been demonstrated that type-I Dirac electron, which has no Fermi pockets, can be realized under ambient pressure \cite{KitouSawaTsumuraya}. 
The calculation considering spin-orbit interaction (SOI) by the second-order perturbation indicated that SOI also contributed to the electronic state in $\alpha$-(BETS)$_2$I$_3$ owing to the presence of \tk{Selenium (Se)}, and its magnitude was $5\sim 10$ meV~\cite{Winter}. 
The results of a recent first-principles calculation with the generalized gradient approximation (GGA) also showed that the SOI had a value of approximately 2 meV, and its effect could not be neglected \cite{TsumurayaSuzumura}. 


In this \Erase{paper}\editage{study}, we \Erase{study}\editage{investigate} the effects of the Coulomb interaction and SOI as \Erase{candidates for the}\editage{possible} cause\editage{s} of the hidden phase transition and insulating behavior at low $T$s. 
We investigate the electronic state and calculate several transport coefficients in $\alpha$-(BETS)$_2$I$_3$. 
The remainder of this paper is organized as follows. 
\tk{In Sec. II,} first-principles calculations based on the X-ray data are performed to derive the transfer integrals at 30 K under ambient pressure. 
\tk{We obtain the on-site Coulomb interaction by the constrained random phase approximation.} 
Using the obtained data, \tk{we construct a  two-dimensional effective Hubbard model}. 
In addition, we \Erase{show}\editage{describe} a method to calculate the DC and optical conductivities using the Nakano-Kubo formula. 
In Sec. III, we demonstrate the \Erase{results of the}\editage{obtained} electronic state at a finite $T$ and a candidate \Erase{for the }low $T$ insulator phase. 
Moreover, a calculation considering SOI is performed, and its contribution to the electronic state near the phase transition is estimated. 
Next, we calculate the $T$-dependence of the DC and optical conductivities \cite{Streda, Shon, Proskurin, Ruegg, Omori2017}. 
$T$- and in-plane magnetic field $B$-dependence of the DC resistivity are also calculated and compared with the experimental results. 
The findings of our study are summarized in Section IV. 
\section{Model and Formulation}

%
\subsection{Effective model based on first-principles calculations}
%

\Erase{As a first step}\editage{First}, we performed first-principles calculations based on the X-ray crystal structural data of $\alpha$-(BETS)$_2$I$_3$ at 30K under ambient pressure \cite{KitouSawaTsumuraya} using the Quantum Espresso (QE) package \cite{Giannozzi}. 
\tk{In our calculation, the GGA was used as the exchange-correlation function \cite{Perdew}. 
As the pseudo-potentials, we used the SG15 Optimized Norm-Conserving Vanderbilt (ONCV) pseudo-potentials \cite{Schlipf}. 
The cutoff kinetic energies for wave functions and charge densities were set as 80 and 320 Ry, respectively. 
The mesh of the wavenumbers was set as $4 \times 4 \times 2$.} 
After the first principles calculation, the maximally localized Wannier functions (MLWFs) were obtained using RESPACK \cite{Nakamura}. 
To construct the MLWFs, four bands near the Fermi energy were selected. 
Initial coordinates of the MLWFs were located at the center of each BETS molecule in the unit cell. 

%
\begin{figure}
\begin{centering}
\includegraphics[width=75mm]{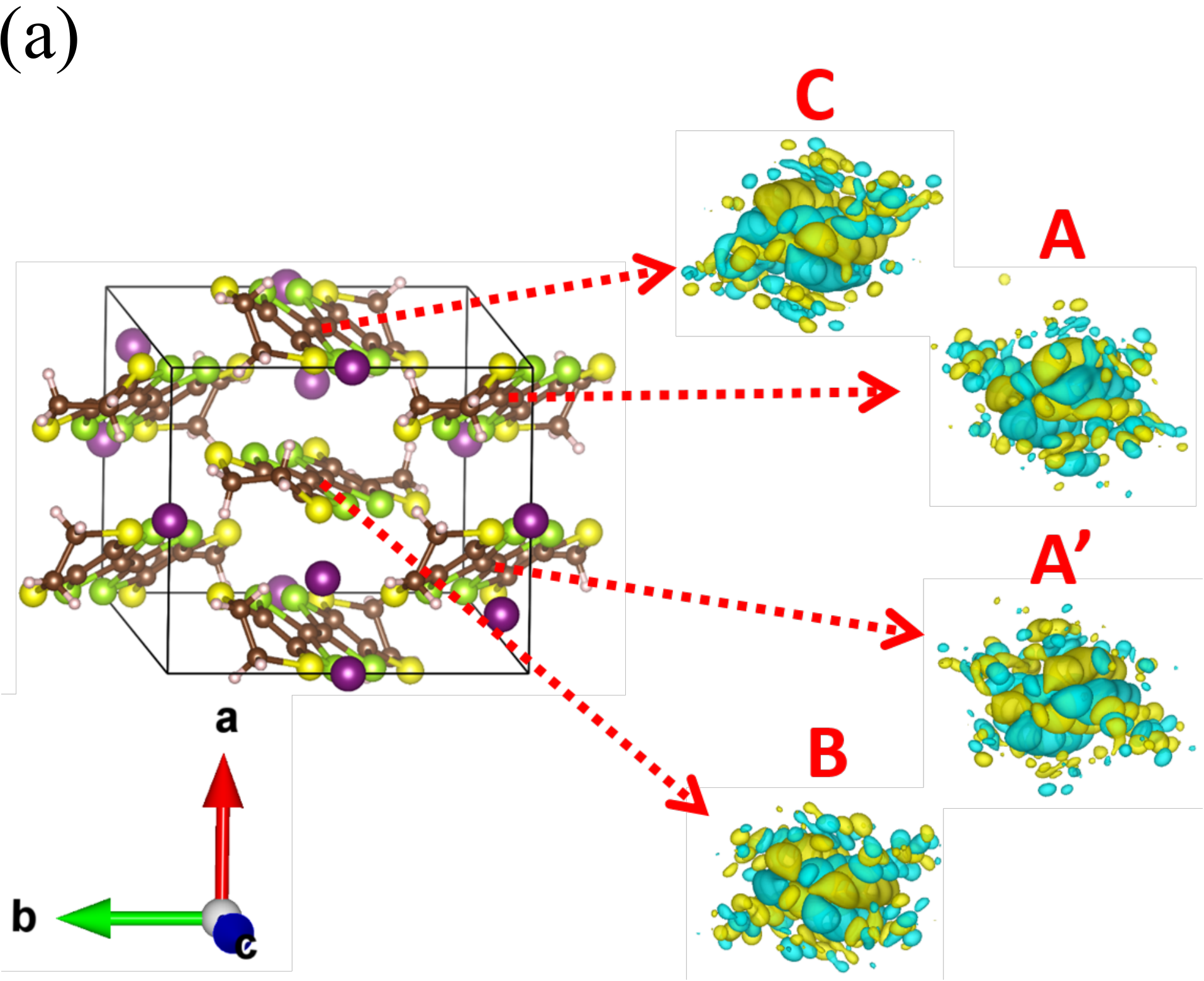}
\includegraphics[width=75mm]{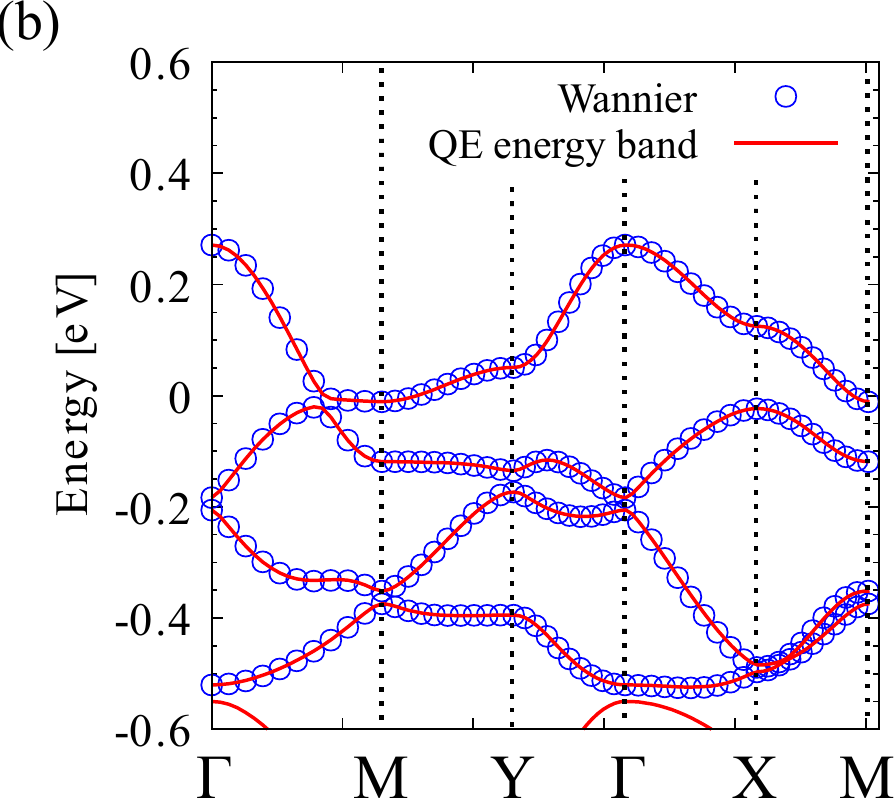}
\caption{\label{illustration}(Color online) 
(a) Crystal structure of $\alpha$-(BETS)$_2$I$_3$ at 30K \tk{under} ambient pressure (left) and real space distribution of the MLWFs (right) \tk{drawn by VESTA \cite{Momma}}. 
(b) Energy bands derived from the first-principles calculation (solid red line) and Wannier interpolation (empty circle). 
The chemical potential is set as the energy origin. 
}\label{Fig:Wannier-Fit}
\end{centering}
\end{figure}
%
Figure \ref{Fig:Wannier-Fit}(a) shows the crystal structure of $\alpha$-(BETS)$_2 $I$_3$ at 30K under ambient pressure (left side) and the real space structure of the MLWFs at each site \editage{(right side)}. 
There are four BETS molecules labeled by A, A$'$, B, and C in the unit cell\editage{.} 
\editage{They} are distinguished by the arrangement and the orientation\editage{.} 
A and A$'$ are crystallographically equivalent sites. 
The center \tk{positions} of the MLWFs are located at the center of \tk{each} BETS molecule, and as shown in Fig. \ref{Fig:Wannier-Fit}(a), $p_z$ like orbitals are spreading \tk{in the direction} perpendicular to the surface of the molecule. 
Figure \ref{Fig:Wannier-Fit}(b) shows the energy bands near the Fermi energy (the energy origin is set as the Fermi energy) obtained by QE and \editage{the Wannier interpolation}. 

%
%
Next, we constructed the effective model using the transfer integrals and \tk{the on-site} Coulomb interactions. 
The on-site Coulomb interactions are \tk{evaluated by the constrained random phase approximation (cRPA) method} using RESPACK. 
The energy cutoff for the dielectric function was set as 5.0 Ry. 
%
\begin{figure*}[tb]
 \begin{tabular}{ll}
   \begin{minipage}{0.55\textwidth}
\begin{centering}
\includegraphics[width=115mm]{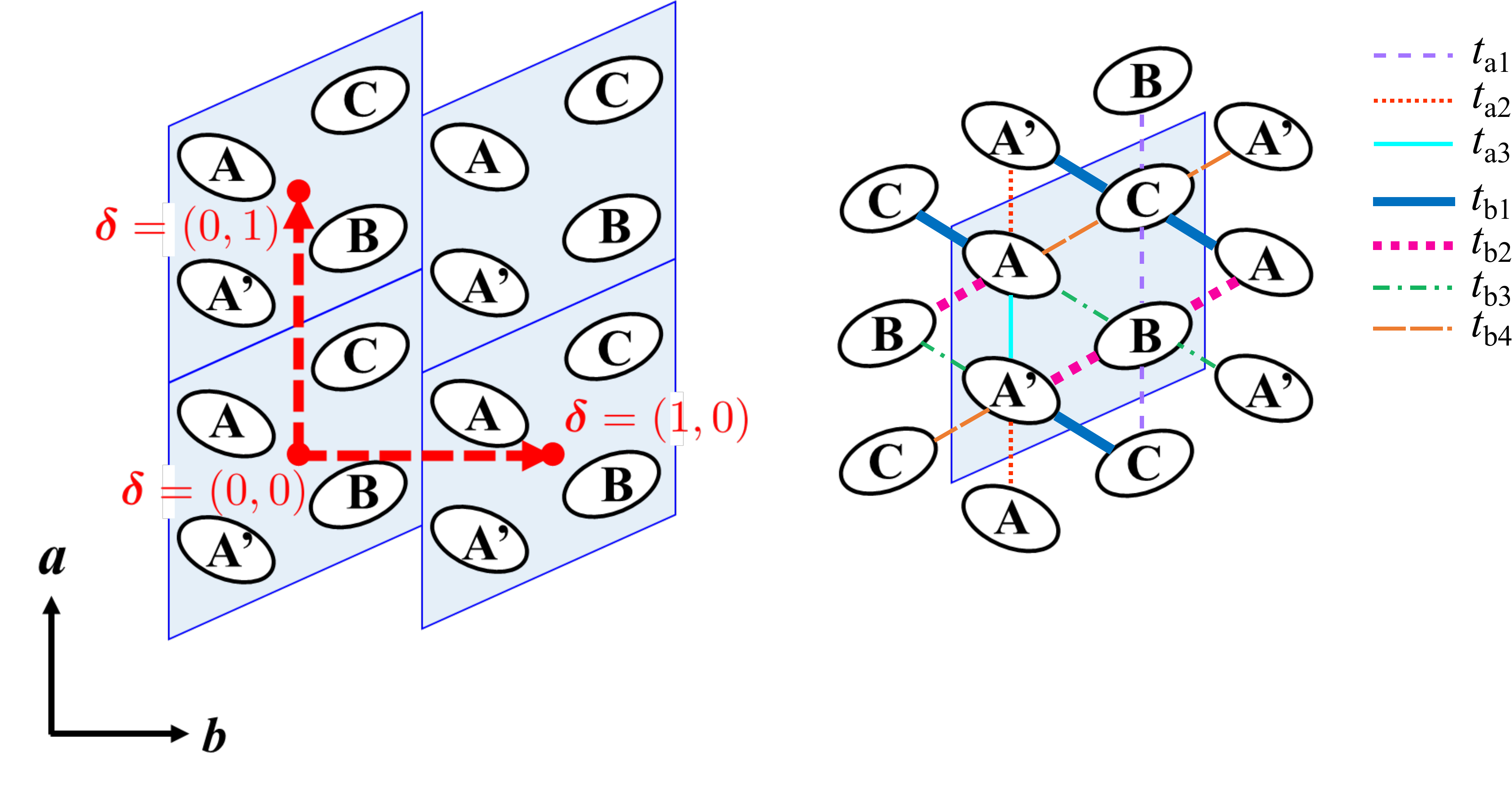}
\end{centering}
\end{minipage}

\begin{minipage}{0.5\textwidth}
 \centering
   {List of transfer integrals obtained \tk{from MLWFs}.}
  \begin{tabular}{llll}
   \hline
   \rule[-3.7mm]{0mm}{8mm}
    ${\bm \delta}=(\delta_b, \delta_a)$ &  $\alpha$ & $\beta$ & Re $\left[ t_{\alpha,\beta}^{{\bm \delta}}\right]$ [meV] \\
    \hline \hline
    (-1, 0) & A & B &  158.7  \hspace{0.3cm}($t_{\rm b2}$)\\ 
    (0, 0) & A$'$ & B &  158.6  \hspace{0.3cm}($t_{\rm b2}$)\\
   (0, -1) & A$'$ & C &  138.1  \hspace{0.3cm}($t_{\rm b1}$)\\
    (-1, 0) & A & C &  138.0  \hspace{0.3cm}($t_{\rm b1}$)\\
    (0, 0) & A & B &  65.84  \hspace{0.3cm}($t_{\rm b3}$)\\
    (-1, 0) & A$'$ & B &  65.75  \hspace{0.3cm}($t_{\rm b3}$)\\ 
    (0, 0) & A & A$'$ &  51.08  \hspace{0.3cm}($t_{\rm a3}$)\\
   (0, -1) & C & C &  21.92  \hspace{0.3cm}($t_{\rm a4}'$)\\
   (-1,-1) & A$'$ & C &  18.65  \hspace{0.3cm}($t_{\rm b4}$)\\ 
    (0, 0) & A & C &  18.48  \hspace{0.3cm}($t_{\rm b4}$)\\
   (0, -1) & A$'$ & A & -16.31  \hspace{0.3cm}($t_{\rm a2}$)\\
   (0, -1) & A$'$ & A$'$ &  14.24  \hspace{0.3cm}($t_{\rm a1}'$)\\
   (0, -1) & A & A &  14.19  \hspace{0.3cm}($t_{\rm a1}'$)\\
   (0, -1) & B & C &  10.12  \hspace{0.3cm}($t_{\rm a1}$)\\ 
    (0, 0) & B & C &  9.864  \hspace{0.3cm}($t_{\rm a1}$)\\
   (-1,-1) & A$'$ & A &  9.212\\ 
   (0, -1) & A$'$ & B &  6.737\\ 
   (1, -1) & B & A &  6.600\\ 
   (1, -1) & B & C &  5.065\\ 
    (-1, 0) & B & C &  5.010\\ 
   \hline
  \end{tabular}
\end{minipage}

 \end{tabular}
\caption{\label{illustration}(Color online) Schematic lattice structure of $\alpha$-(BETS)$_2$I$_3$. 
The area of the original unit cell is shown by the shaded blue region.
\tk{The dotted red arrows in left figure indicate the relative lattice vectors ${\bm \delta}=(\delta_b,\delta_a)$ for ${\bm \delta}=(1,0)$ and $(0,1)$, as an examples.
The center figure shows the transfer integrals between nearest neighbor sites from the original unit cell.}\label{Fig:2Dnetwork-TransferTable}
}
\end{figure*}

%
Figure \ref{Fig:2Dnetwork-TransferTable} shows a schematic lattice structure of $\alpha$-(BETS)$_2$I$_3$. 
The transfer integrals are considered up to almost the next nearest neighbor bonds \Erase{(enclosed by the red broken line in Fig. \ref{Fig:2Dnetwork-TransferTable})}\tk{(shown in the center figure and table of Fig. \ref{Fig:2Dnetwork-TransferTable})}.
The values of the transfer integrals $t_{\alpha,\beta}^{{\bm \delta}}$ are listed in the table shown in the right side of Fig. \ref{Fig:2Dnetwork-TransferTable}. 
Here, ${\bm \delta} = (\delta_b, \delta_a)$ indicates the \tk{relative }lattice vector and $\alpha$ and $\beta$ indicate the site indexes in the unit cell, i.e., A, A', B, and C. 
The cutoff energy of the transfer integrals are taken as $t_{\rm cut} = 5.0$ [meV]. 
The on-site Coulomb interactions are given as $U_{\rm A}=U_{{\rm A}'}=1.383$ [eV], $U_{\rm B}=1.396$ [eV], and $U_{\rm C}=1.359$ [eV]. 
Since the transfer integrals between the inter planes are \Erase{much}\editage{significantly} smaller than those in the intra plane \cite{InterPlaneTransfers}, this system can be considered as a two-dimensional electron system. 

In this \Erase{paper}\editage{study}, we investigated the two-dimensional Hubbard model \tk{with SOI \cite{KinoFukuyama1995, KinoFukuyama1996}}:  
\tk{
\begin{eqnarray}
H&=&\sum_{\bm{R}, \bm{\delta}}\sum_{\alpha,\beta}{\sum_{{\sigma}}t^{\bm \delta}_{{\alpha},\beta}}c^{\dag}_{{\bm R},{\alpha},{\sigma}}c_{{\bm R}+\bm{\delta},\beta,{\sigma}}+{\sum_{{\bm R},{\alpha}}}\lambda_U U_{\alpha}n_{{\bm R},{\alpha},{\uparrow}}n_{{\bm R},{\alpha},{\downarrow}}\nonumber\\
&&+H^{\rm SOI}-\mu_{\rm B}B\sum_{\alpha,\sigma,{\bm R}}{\rm sgn}(\sigma)n_{{\bm R},{\alpha},{\sigma}},
\end{eqnarray}
}
where $\bm R$ is the coordinate of the unit cell, and $\alpha$, $\beta$ indicate the indexes of the inner-sites in the unit cell (A, A$'$, B, and C). 
$\sigma = \uparrow (+), \downarrow (-)$ indicates the index of spin. 
\tk{
$t_{\alpha,\beta}^{\bm \delta}$ indicates the transfer integral between $\alpha$ and $\beta$ sites separated by the relative lattice vector ${\bm \delta}$, and $U_\alpha$ indicates the on-site Coulomb interaction evaluated using cRPA method.}
\tk{
Here, the site potentials $t_{\alpha}^{\bm 0}$ are $t_{\rm A}^{\bm 0} = t_{\rm A'}^{\bm 0} = 4.467$ [eV], $t_{\rm B}^{\bm 0} = 4.462$ [eV], and $t_{\rm C}^{\bm 0} = 4.475$ [eV]. 
We ignored these terms in eq. (1) because their contribution to the energy band obtained in our model is insignificant.
}
The creation (annihilation) operator at $\alpha$-site in the unit cell located at $\bm R$ is defined as $c_{{\bm R},{\alpha},{\sigma}}$ ($c^{\dag}_{{\bm R},{\alpha},{\sigma}}$), and the number operator is defined as $n_{{\bm R},\alpha,\sigma} = c^{\dag}_{{\bm R},{\alpha},{\sigma}}c_{{\bm R},\alpha,{\sigma}}$. 
$\lambda_U$ ($0 < \lambda_U < 1$) is a \tk{tuning} parameter \Erase{which}\editage{that} control\tk{s} the value\tk{s} of \tk{the on-site} Coulomb interaction. 
$H^{\rm SOI}$ is the SOI term, which is generally proportional to $\left({{\bf p}\times\nabla U({\bf r})}\right)\cdot\bm{\sigma}$, where $\bf p$ is the momentum, $U({\bf r})$ is the potential energy, and $\bm \sigma$ \Erase{means}\editage{indicates} the spin angular momentum. 
\tk{The specific formula of $H^{\rm SOI}$ is detailed in the following section.} 
The fourth term of Eq. (1) represents the in-plane Zeeman magnetic field, where $\mu_{\rm B}$ is the Bohr magneton. 
In the following, the lattice constants, Boltzmann constant $k_B$, and the Plank constant $\hbar$ are taken as unity. 
Note that electronvolt (eV) is used as the unit of energy throughout this paper.

\subsection{Electronic state in the wavenumber space}
In this \Erase{paper}\editage{study}, we \Erase{study}\editage{investigate} the electronic state using the Hartree approximation. 
\Erase{For}\editage{To} obtain\Erase{ing} the Hamiltonian in the wavenumber representation, the Fourier inverse transformation is performed \Erase{to}\editage{on} the Hamiltonian defined in Eq. (1). 
Then, the Hamiltonian is given as
%
\begin{eqnarray}
H_{\alpha,\beta,\sigma}({\bf k}) &=& \sum_{\bm{\delta}}t_{\alpha,\beta}^{(\bm{\delta})}e^{i{\bf k}\cdot\bm{\delta}}c^\dag_{{\bf k},\alpha,\sigma}c_{{\bf k},\beta,\sigma}\nonumber\\
&&+\delta_{\alpha\beta}\lambda_UU_\alpha\langle n_{\alpha,-\sigma}\rangle c^\dag_{{\bf k},\alpha, \sigma}c_{{\bf k},\alpha,\sigma}\nonumber\\
&&+H^{\rm SOI}_{\alpha,\beta,\sigma}({\bf k})\nonumber\\
&&-\mu_{\rm B}B\sum_{\alpha,\sigma,{\bf k}}{\rm sgn}(\sigma) c^\dag_{{\bf k},\alpha, \sigma}c_{{\bf k},\alpha,\sigma},
\end{eqnarray}
where ${\bf k} = (k_b, k_a)$ indicates the wavenumber \tk{vector}. 
\tk{Here,} $H^{\rm SOI}_{\alpha\beta\sigma}({\bf k})$ \tk{is the Hamiltonian of the SOI and is given as the following formulas \cite{Osada}:}
%
\begin{eqnarray}
H^{\rm SOI}_{{\rm B, A},\sigma}({\bf k})&=&i\lambda_{\rm SOI}S_z\left(-t_{\rm B,A}^{(0,0)}+t_{\rm B,A}^{(1,0)}e^{ik_b}\right)c^\dag_{{\bf k},{\rm B},\sigma}c_{{\bf k},{\rm A},\sigma},\nonumber\\
H^{\rm SOI}_{{\rm B,A'},\sigma}({\bf k})&=&i\lambda_{\rm SOI}S_z\left(t_{\rm B,A'}^{(0,0)}-t_{\rm B,A'}^{(1,0)}e^{ik_b}\right)c^\dag_{{\bf k},{\rm B},\sigma}c_{{\bf k},{\rm A'},\sigma},\nonumber\\
H^{\rm SOI}_{{\rm C,A},\sigma}({\bf k})&=&i\lambda_{\rm SOI}S_z\left(-t_{\rm C,A}^{(0,0)}+t_{\rm C,A}^{(1,0)}e^{ik_b}\right)c^\dag_{{\bf k},{\rm C},\sigma}c_{{\bf k},{\rm A},\sigma},\nonumber\\
H^{\rm SOI}_{{\rm C,A'},\sigma}({\bf k})&=&i\lambda_{\rm SOI}S_z\left(t_{\rm C,A'}^{(0,1)}e^{ik_b}-t_{\rm C,A'}^{(1,1)}e^{i(k_b+k_a)}\right)\nonumber\\
&&\times c^\dag_{{\bf k},{\rm C},\sigma}c_{{\bf k},{\rm A'},\sigma}, \nonumber
\end{eqnarray}
where the spin $S_z = {\rm sgn}(\sigma)/2$ and $\lambda_{\rm SOI}$ is the control parameter of the strength of the SOI.
\Erase{
In the present study, $\lambda_{\rm SOI}$ and $S_z$ are treated as constants for simplicity, since the results shown in the following sections can be explained well in the range of this approximation and robust for any other treatment of SOI \cite{Ezawa, Mong, Zheng, Miyakoshi, Cao, Hohenadler, Sekine, Jiang}.
However, more exactly, it is necessary to treat these as vector quantities in consideration of the anisotropy of SOI \cite{Winter}.
}

\tk{$H_{\alpha,\beta,\sigma}({\bf k})$} is diagonalized by using the eigenvector $d_{\alpha, \nu, \sigma}({\bf k})$ about each $\bf k$, and the energy eigenvalues
\tk{$\tilde{E}_{\nu, \sigma}({\bf k})=\bigl{\langle}\sum_{\alpha, \beta} d^*_{\alpha, \nu, \sigma}({\bf k}) H_{\alpha, \beta, \sigma}({\bf k}) d_{\beta, \nu, \sigma}({\bf k})\bigr{\rangle}$ ($\tilde{E}_{1, \sigma}({\bf k}) > \tilde{E}_{2, \sigma}({\bf k}) > \tilde{E}_{3, \sigma}({\bf k}) > \tilde{E}_{4, \sigma}({\bf k})$) are obtained.} 
\tk{In the following, for convenience, we define $E_{\nu, \sigma}(\bf k)$ as}
%
\begin{eqnarray}
E_{\nu, \sigma}({\bf k})=\biggl{\langle}\sum_{\alpha, \beta} d^*_{\alpha, \nu, \sigma}({\bf k}) H_{\alpha, \beta, \sigma}({\bf k}) d_{\beta, \nu, \sigma}({\bf k})\biggr{\rangle}-\mu, 
\end{eqnarray}
where the chemical potential $\mu$ is determined to satisfy the \tk{\Erase{$\frac{3}{4}$} $3/4$-}filling. 
The charge density $\langle n_{\alpha, \sigma}\rangle$ for site $\alpha$ and spin $\sigma$ is calculated as $\langle n_{\alpha, \sigma} \rangle = \sum_{{\bf k}, \nu}|d_{\alpha, \nu, \sigma}({\bf k})|^2f(E_{\nu, \sigma}({\bf k}))$ using the Fermi distribution function \tk{$f(\xi) = \left[ 1 + \exp(\xi/T) \right]^{-1}$ \Erase{$f(E_{\nu,\sigma}({\bf k})) = \left[ 1 + \exp(E_{\nu,\sigma}({\bf k})/T) \right]^{-1}$}}. 
The Berry curvature $B_{\nu,\sigma}({\bf k})$ in band $\nu$ and spin $\sigma$ is obtained by
%
\begin{eqnarray}
B_{\nu,\sigma}({\bf k}) = \sum_{\nu'\neq\nu}\frac{v^b_{\nu,\nu',\sigma}({\bf k})v^a_{\nu',\nu,\sigma}({\bf k})}{i(E_{\nu,\sigma}({\bf k})-E_{\nu',\sigma}({\bf k}))^2}+{\rm c.c.},
\end{eqnarray}
where
\begin{eqnarray}
v^\gamma_{\nu,\nu',\sigma}({\bf k}) = \sum_{\alpha,\beta}d^*_{\alpha,\nu,\sigma}({\bf k})\frac{\partial H_{\alpha,\beta,\sigma}({\bf k})}{\partial k_\gamma}d_{\beta,\nu',\sigma}({\bf k}),
\end{eqnarray}
and the Chern number is given \tk{as \Erase{by the follows}}. 
\begin{eqnarray}
Ch = \sum_\sigma Ch_\sigma = \frac{1}{2\pi}\sum_\sigma\int_{BZ}d{\bf k}B_{\nu,\sigma}({\bf k}).
\end{eqnarray}
Here, $\int_{BZ}$ \Erase{means}\editage{indicates} that the \Erase{integral}\editage{integration} is performed throughout the Brillouin zone. 

\subsection{Conductivity}
The optical conductivity in the clean limit is calculated using the Nakano-Kubo formula \cite{Streda, Shon, Proskurin, Ruegg, Omori2017} given as follows
\begin{eqnarray}
\sigma(\omega, \theta)&=&\frac{1}{i\omega}\left[ Q^R(\omega, \theta)-Q^R(0, \theta) \right],\\
Q^R(\omega, \theta)&=&\frac{e^2}{N_L}\sum_{{\bf k}, \nu, \nu', \sigma} |{\rm v}_{\nu,\nu',\sigma}({\bf k}, \theta)|^2\nonumber\\
&&\times\chi_{\nu,\nu',\sigma}^0({\bf k},\omega),\\
\chi_{\nu,\nu',\sigma}^0({\bf k},\omega)&=&-\frac{f(E_{\nu,\sigma}({\bf k}))-f(E_{\nu',\sigma}({\bf k}))}{E_{\nu,\sigma} ({\bf k})-E_{\nu',\sigma}({\bf k})+\hbar \omega+i0^+}, 
\end{eqnarray}
where $0^+=5.0 \times 10^{-4}$ and the angle $\theta$ is measured from the $b$-axis direction and the projection in the $\theta$\tk{-}direction of the velocity ${\rm v}_{\nu,\nu',\sigma}({\bf k}, \theta)$ indicating the inter\tk{-}band transition\Erase{ is} written as 
\begin{eqnarray}
{\rm v}_{\nu,\nu',\sigma}({\bf k}, \theta) = \sum_{\alpha, \beta}d^*_{\alpha,\nu,\sigma}({\bf k})v_{\alpha,\beta,\sigma}({\bf k}, \theta)d_{\beta,\nu',\sigma}({\bf k}).
\end{eqnarray}
Here\tk{,} $v_{\alpha,\beta,\sigma}({\bf k},\theta)$ is defined as 
\begin{eqnarray}
v_{\alpha,\beta,\sigma}({\bf k},\theta)&=&\frac{1}{\hbar}\biggl(\frac{\partial H_{\alpha,\beta,\sigma}({\bf k})}{\partial {k_x}}\cos\theta\nonumber\\
&&+\frac{\partial H_{\alpha,\beta,\sigma}({\bf k})}{\partial {k_y}}\sin\theta\biggr). 
\end{eqnarray}

In the limit of $\omega\to0$ in Eq. (7), the DC conductivity is represented by the following equations: 
\begin{eqnarray}
\sigma(\theta)&=&\int d\omega \left(-\frac{df}{d\omega} \right)\Phi(\omega,\theta),\\
\Phi(\omega,\theta)&=&\frac{2 e^2}{N_L}\sum_{{\bf k},\nu,\sigma}\left|{\rm v}_{\nu,\sigma}({\bf k}, \theta)\right|^2\tau_{\nu,\sigma}(\omega,{\bf k})\nonumber\\
&&\times\delta(\hbar \omega-E_{\nu,\sigma}({\bf k})),
\end{eqnarray}
where the relaxation time $\tau_{\nu,\sigma}(\omega,{\bf k})$ \tk{is calculated} within the $T$-matrix approximation using the perturbation theory \tk{for} the green function. 
\tk{We only treat an elastic scattering between electrons and impurities, which is originated from a lack and disorder of anion I$_3^-$ molecules.}
The impurity potential term is considered as
\begin{eqnarray}
H_{\rm imp}=\frac{V_0}{N_L}\sum_{{\bf k},{\bf q},\alpha,\sigma}^{\rm imp}\sum_{i}e^{-i{\bf q}\cdot {\bf r}_i}c^{\dag}_{{\bf k}+{\bf q},\alpha,\sigma}c_{{\bf k},\alpha,\sigma},
\end{eqnarray}
where $V_0$ is the intensity of the impurity potential and $ {\bf r}_i$ is the coordinate of impurities. 
The imaginary part of the retarded self-energy ${\rm Im}\Sigma^R_{\nu,\sigma}(\omega,{\bf k})$ gives the damping constant $\gamma_{\nu,\sigma}(\omega,{\bf k})$ and the $\tau_{\nu,\sigma}(\omega,{\bf k})$ is obtained as follows. 
\begin{eqnarray}
\gamma_{\nu,\sigma}(\omega,{\bf k})&=&\frac{\hbar}{2\tau_{\nu,\sigma}(\omega,{\bf k})}=-{\rm Im}\Sigma^R_{\nu,\sigma}(\omega,{\bf k})\nonumber\\
&=&c_{\rm imp}\frac{ \left|d_{\alpha,\nu,\sigma}({\bf k})\right|^2\left\{ \pi V_0^2 \mathcal{N}_\sigma(\omega)\right\}}{1+\left\{ \pi V_0 \mathcal{N}_\sigma(\omega)\right\}^2}.
\end{eqnarray}
Here, $c_{\rm imp} \ll 1$ is the density of impurities and 
$$\mathcal{N}_\sigma(\omega) = \sum_{{\bf k},\alpha,\nu}|d_{\alpha, \nu, \sigma}({\bf k})|^2\delta(\hbar\omega-E_{\nu,\sigma}({\bf k}))$$
indicates the total density of states. 
In the following, the unit of conductivity \Erase{means}\editage{is} the universal conductivity $\sigma_0=4e^2/\pi h$, and the Drude term is subtracted from the optical conductivity.

%
\section{Numerical Results}
%

%
\subsection{Electronic state at finite temperature}
In this subsection, the electronic state at a finite $T$ is investigated under the condition of $\lambda_{\rm SOI} = 0$. 
%
\begin{figure}
\begin{centering}
\includegraphics[width=60mm]{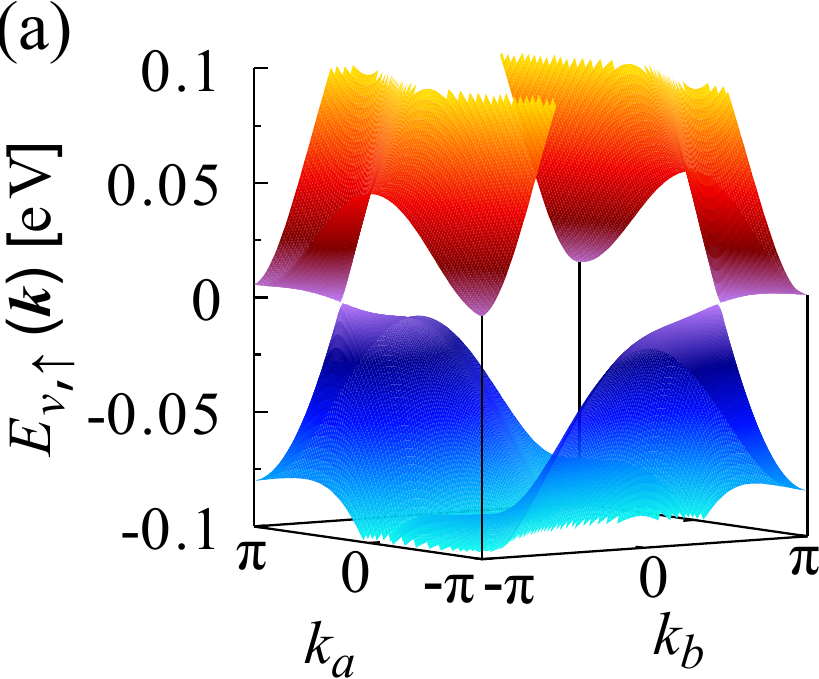}
\includegraphics[width=60mm]{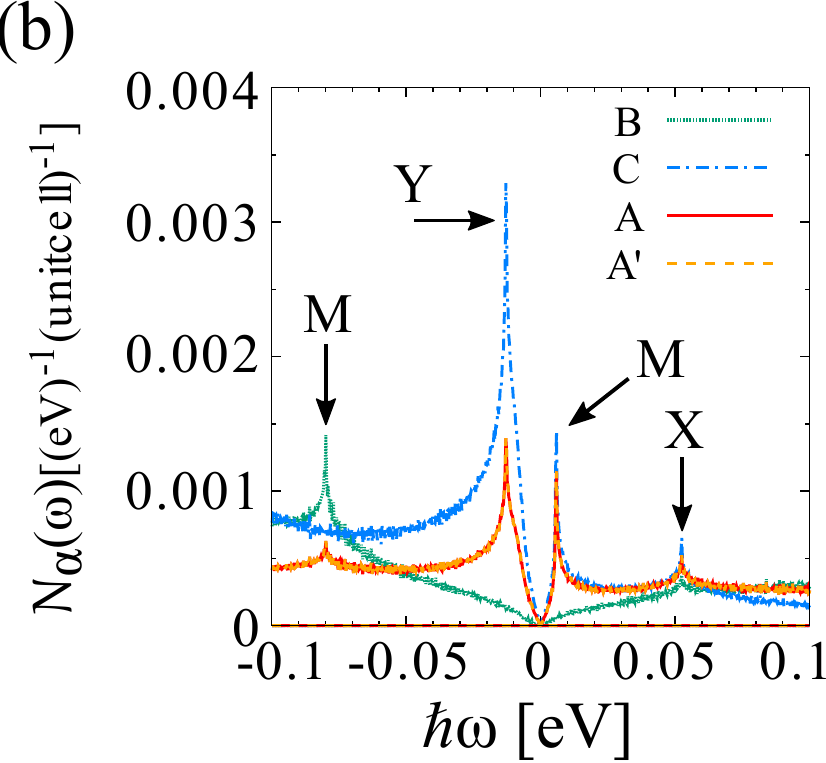}
\includegraphics[width=60mm]{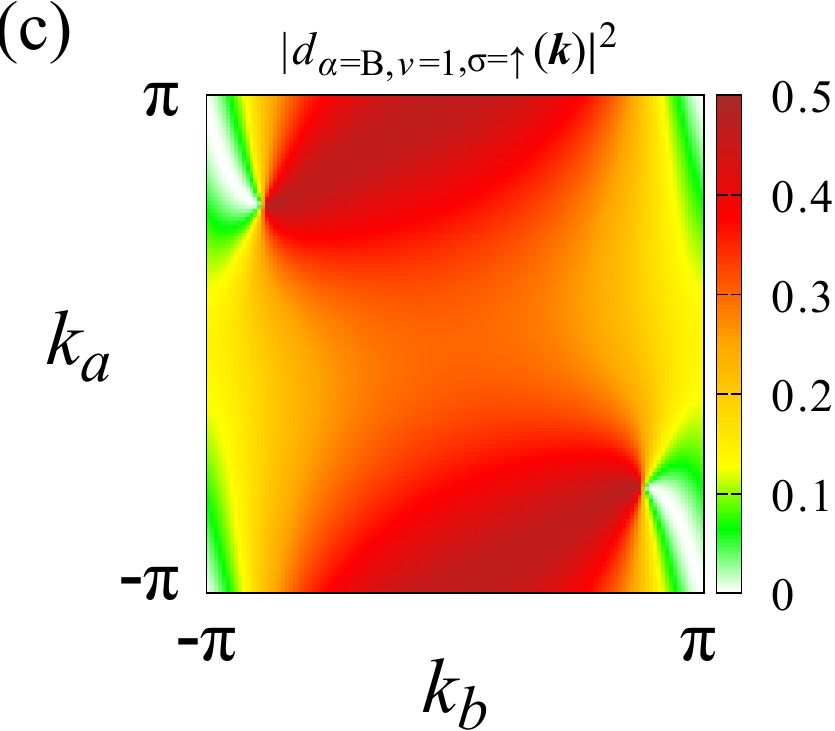}
\includegraphics[width=60mm]{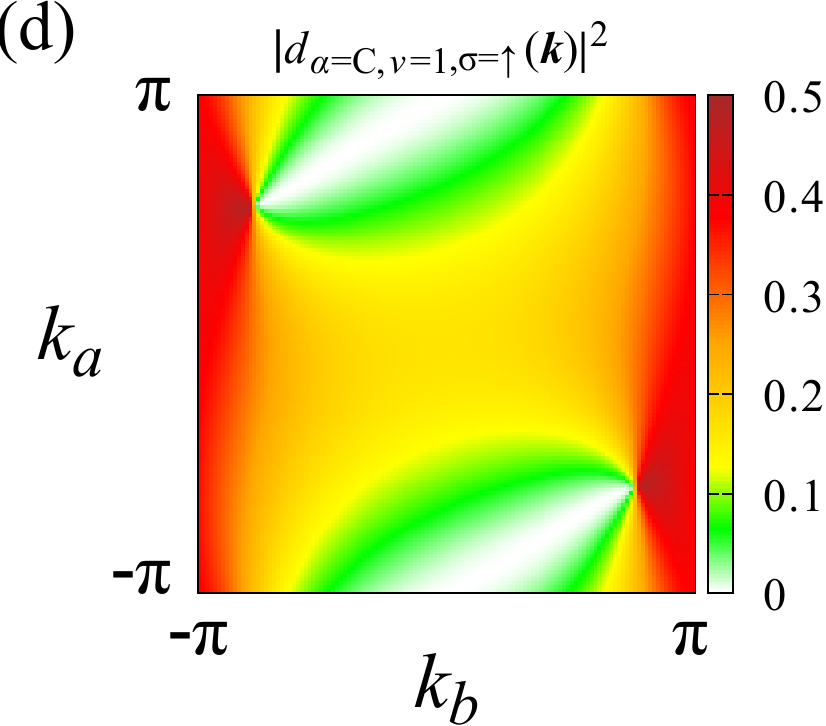}
\caption{\label{illustration}(Color online) (a) Energy eigenvalues $E_{\nu,\sigma=\uparrow}({\bf k})$ ($\nu = 1, 2$) calculated \tk{on the basis of} the tight-binding model, (b) $\mathcal{N}_\alpha (\omega)$, and square of the absolute value of eigenvectors $\left| d_{\alpha,\nu = 1,\sigma = \uparrow}({\bf k})\right|^2$ at (c) $\alpha = $ B and (d) $\alpha =$ C. 
The symbols of X, Y, and M in (b) indicate the symmetric points in the Brillouin zone corresponding to the van \tk{Hove} singularity. 
}\label{Fig:TBResult}
\end{centering}
\end{figure}
%
Figure \ref{Fig:TBResult}(a) shows the energy eigenvalues $E_{\nu,\sigma}({\bf k})$ near the Fermi energy calculated \Erase{by}\editage{using} the tight-binding model ($\lambda_{U} = 0$). 
The conduction band ($\nu = 1$) and valence band ($\nu = 2$) form the Dirac point, and a type-I Dirac electron system that appears in the high-pressure phase of $\alpha$-(BEDT-TTF)$_2$I$_3$ is expected to be realized under ambient pressure in $\alpha$-(BETS)$_2$I$_3$. 

Figure \ref{Fig:TBResult} (b) displays the density of states $\mathcal{N}_{\alpha}(\omega)$ \tk{in the energy range $|\hbar\omega| < 0.1$}.
The order of $\mathcal{N}_{\alpha}(\omega)$ magnitudes near the Fermi energy (\Erase{$\hbar \omega = 0$}\tk{approximately $|\hbar\omega| < 0.05$}) is $\mathcal{N}_{\rm C}(\omega) > \mathcal{N}_{\rm A}(\omega) \tk{= \mathcal{N}_{\rm A'}(\omega)} > \mathcal{N}_{\rm B}(\omega)$.
The presence or absence of peaks of the van \tk{Hove} singularity at each site is related to the property of the eigenvector $d_{\alpha,\nu,\sigma}({\bf k})$.
\tk{
Note that the lines of $N_{\rm A}(\omega)$ and $N_{\rm A’}(\omega)$ has the same value due to the inversion symmetry and overlap each other.
}
Figure \ref{Fig:TBResult}(c) and (d) show the square of the absolute value of the eigenvector $\left| d_{\alpha,\nu = 1,\sigma = \uparrow}({\bf k})\right|^2$ in $\alpha = $ B and C, respectively. 
The zero line appears in $\left| d_{\alpha,\nu = 1,\sigma = \uparrow}({\bf k})\right|^2$, which has almost the same wavenumber dependence as $\alpha$-(BEDT-TTF)$_2$I$_3$~\cite{Kobayashi2013}. 
Accordingly, the electronic state of $\alpha$-(BETS)$_2$I$_3$ in the high-$T$ phase under ambient pressure \Erase{has an electronic state}\editage{is} similar to this, as demonstrated by $\alpha$-(BEDT-TTF)$_2$I$_3$ in the high-pressure phase. 
%

%
\begin{figure}
\begin{centering}
\includegraphics[width=76mm]{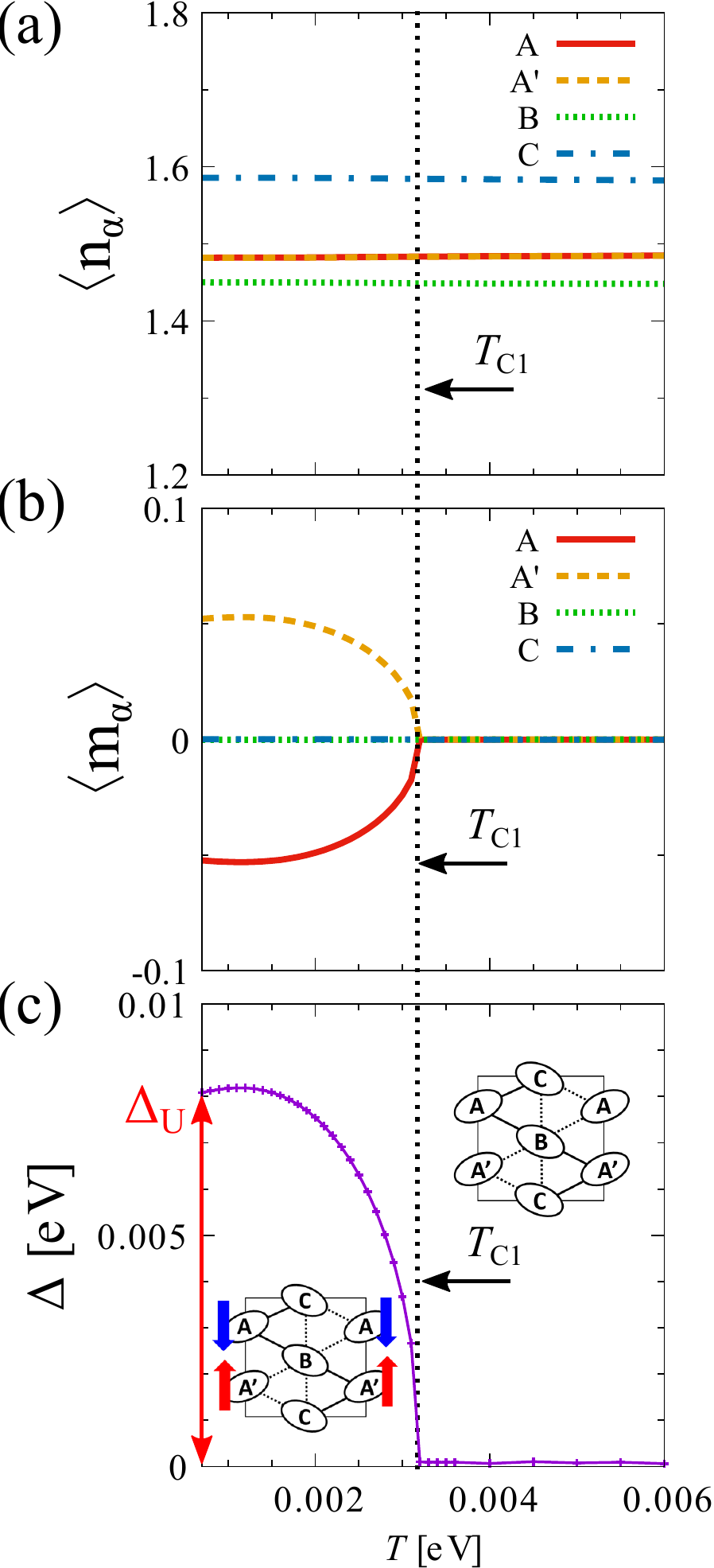}
\caption{\label{illustration}(Color online) $T$-dependence of (a) charge densities $\langle n_\alpha\rangle$, (b) magnetization densities $\langle m_\alpha \rangle$, and (c) energy gap $\Delta$ at $\lambda_U = 0.344$. 
The black \tk{dotted} line \tk{is plotted as a guide\Erase{ line} to show the temperature} $T = T_{\rm C1} = 0.0032$ where the spin-order phase transition occurs. 
Schematic diagrams of the magnetization density in the unit cell at $T > T_{\rm C1}$ and $T < T_{\rm C1}$ are shown in the inset of (c).
\tk{The energy gap at $T=0.0005$ ($\Delta_U$) is shown by red solid line.}
}\label{Fig:ChargeSpinGap}
\end{centering}
\end{figure}
%
\tk{Hereafter, we fixed $\lambda_U$ as $0.344$, so that the phase transition $T$ matches to that observed in the experiments} and investigated the effects of the on-site Coulomb interaction within the Hartree approximation.
Figure \ref{Fig:ChargeSpinGap}(a) and (b) show the $T$-dependence of the charge density $\langle n_\alpha\rangle$ and magnetization density $\langle m_\alpha\rangle$ at each site in the unit cell.
\tk{It is \Erase{seen}\editage{observed} that with decreasing $T$ from $T=0.006$, the charge densities hardly change, \Erase{while}\editage{whereas} the spin densities at A and A$'$ sites change rapidly at the temperature $T_{\rm C1} \simeq 0.0032$.
\tk{
Here, $\langle n_B\rangle$ and $\langle n_C\rangle$ have different values owing to the charge disproportionation originated from the anisotropy of transfer integrals and it is not related to the charge order.
}
These results indicate that the system does not break the charge inversion symmetry, but breaks the spin inversion symmetry below $T_{\rm C1}$.}
\tk{
Such a magnetic phase transition has not been observed so far, but the divergent increase of spin susceptibility associated with this spin order in $T < T_{\rm C1}$ is probably canceled out due to the nature of the wave function in Dirac electron systems.
}
In the previous \tk{theoretical study} \cite{KinoFukuyama1995, KinoFukuyama1996}, antiferromagnetism in the unit cell with vertical-stripe charge order was pointed out. 
\tk{However, the structure analysis \tk{in the experiments} shows that the \tk{charge} inversion symmetry is not broken \tk{and charge density at each site is hardly changed as $T$ is decreased }\Erase{in the low $T$ phase }\cite{KitouSawaTsumuraya}\Erase{, and this result}\tk{.}
\tk{This fact }is consistent with our results.}
\tk{
The remainder of this paper, we theoretically investigate the anomalous competitive effect with $U$ and SOI whether magnetic transition actually occurs or not, when such a spin order exists in $\alpha$-(BETS)$_2$I$_3$.
}
Figure \ref{Fig:ChargeSpinGap}(c) \tk{shows} the $T$-dependence of the energy gap $\Delta$. 
$\Delta$ has a finite value at $T <  T_{\rm C1}$ owing to the occurrence of the spin-order phase transition.

%
\begin{figure}
\begin{centering}
\includegraphics[width=80mm]{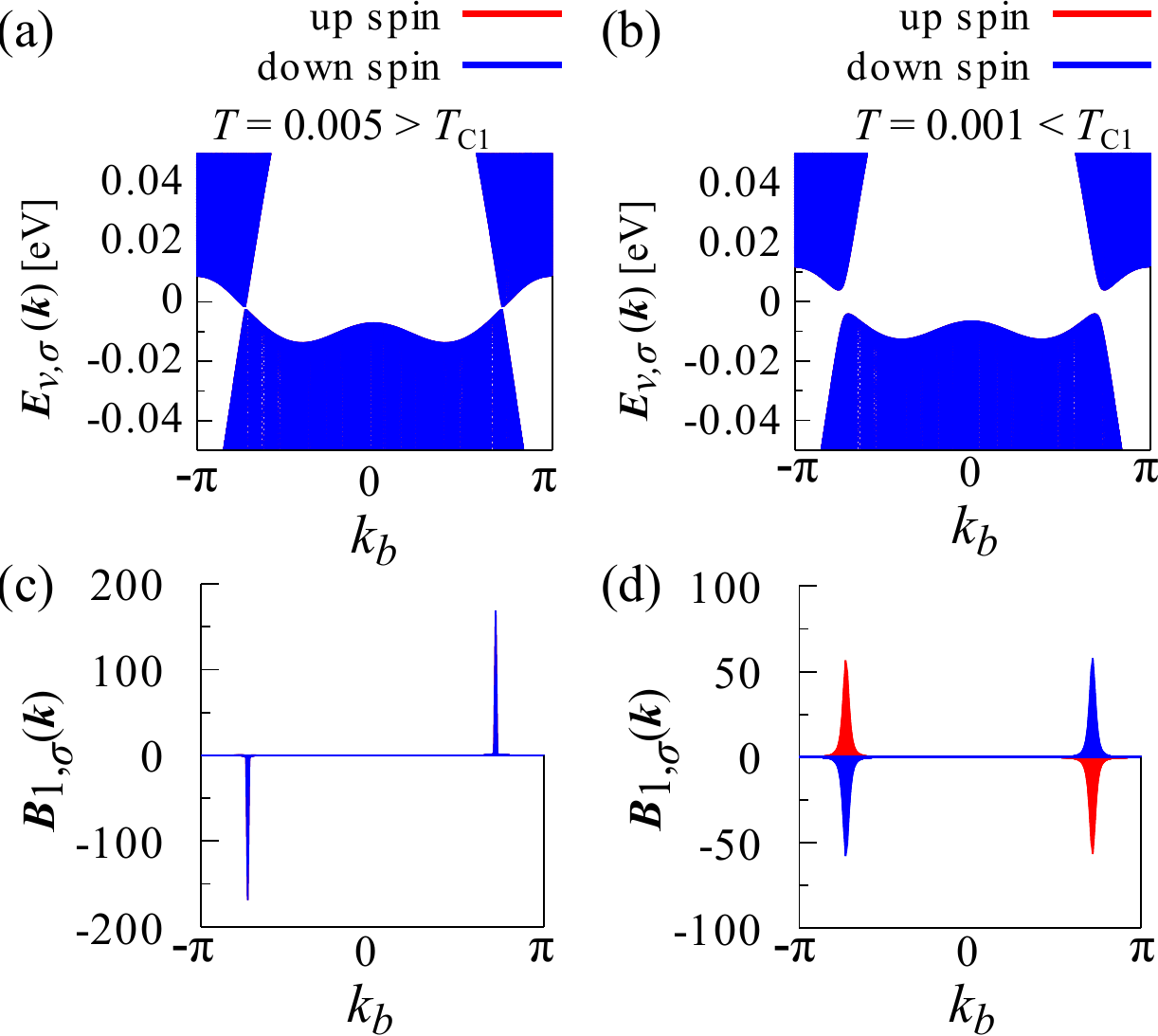}
\caption{\label{illustration}(Color online) Energy eigenvalues $E_{\nu, \sigma}({\bf k})$ for \tk{$\nu=1, 2$ at (a) $T = 0.005 ~(> T_{\rm C1}=0.0032)$ and (b) $T = 0.001 ~(< T_{\rm C1})$, respectively}. 
\tk{Berry curvatures $B_{1,\sigma}({\bf k})$ at (c) $T = 0.005 ~(> T_{\rm C1})$ and (d) $T = 0.001 ~(< T_{\rm C1})$, respectively.} 
}\label{Fig:Band-BC}
\end{centering}
\end{figure}
Figure \ref{Fig:Band-BC}(a) and (b) \tk{show} the energy bands \tk{at $T = 0.005~(> T_{\rm C1})$ and $T = 0.001~(< T_{\rm C1})$, respectively}. 
\tk{In the spin-ordered state,} $\Delta$ opens at the Dirac point, but the spin components of the energy bands do not split. 
On the other hand, Fig\tk{s.} \ref{Fig:Band-BC}(c) and (d) \tk{show} the Berry curvature\tk{s} $B_{1, \sigma}({\bf k})$ at $\sigma = \uparrow$ and $\downarrow$. 
As shown in Fig. \ref{Fig:Band-BC}(c) and (d), the sign of $B_{1, \sigma}({\bf k})$ inverts according to the degrees of freedom about spin $\sigma = \uparrow (+), \downarrow (-)$ and valley indices \tk{$\tau= +1 (-1)$, where the right (left) Dirac cone corresponds to $\tau = +1 (-1)$, respectively.}
Therefore, when such a spin-ordered massive Dirac electron (SMD) phase exists, it is expected that a unique spin-valley Hall effect appears. 
The intrinsic and side-jump terms of the valley-spin Hall conductivity on $\nu$-th band can be written in the form of 
$$\sigma^{\rm H, int}_{\nu,\sigma,\tau} = \frac{e^2}{h}\int d{\bm k}f(E_{\nu,\sigma,\tau}({\bm k}))B_{\nu,\sigma,\tau}({\bm k}),$$
and 
$$\sigma^{\rm H, side}_{\nu,\sigma,\tau} = -\frac{e^2}{h}\int d{\bm k} B_{\nu,\sigma,\tau}({\bm k}) \frac{\partial f(E_{\nu,\sigma,\tau}({\bm k}))}{\partial E_{\nu,\sigma,\tau}({\bm k})}\frac{\partial f(E_{\nu,\sigma,\tau}({\bm k}))}{\partial{\bm k}},$$ 
where $E_{\nu,\sigma,\tau}({\bm k})$ is the energy band at the wavenumber around the left $(\tau=-1)$ or right $(\tau=+1)$ Dirac point \cite{MurakamiNagaosa, Xiao2007, Matsuno2016}. 
The Hall conductivity $\sigma^H$ is defined by $\sigma^{\rm H}_{\nu,\sigma,\tau} = \sigma^{\rm H, int}_{\nu,\sigma,\tau}+\sigma^{\rm H, side}_{\nu,\sigma,\tau}$.
The \tk{spin} and valley Hall conductivities are calculated by $\sigma^{\rm S}_{\nu,\tau} = \sum_\sigma{\rm sgn}(\sigma)\sigma^{\rm H}_{\nu,\sigma,\tau}$ and $\sigma^{\rm V}_{\nu,\sigma} = \sum_\tau{\rm sgn}(\tau)\sigma^{\rm H}_{\nu,\sigma,\tau}$, respectively. 
\Erase{Then}\editage{Subsequently}, the spin-valley Hall conductivity $\sigma^{\rm SV}_{\nu}$ is obtained by $\sigma^{\rm SV}_{\nu}=\sum_{\sigma,\tau}{\rm sgn}(\sigma\tau)\sigma^{\rm H}_{\nu,\sigma,\tau}$ and this value becomes finite in the SMD phase. 
It is expected that the spin (valley) Hall effect depending on the degrees of freedom about valley (spin) appears \cite{Xiao2012}. 
\subsection{Effects of SOI on the electronic state}
In this subsection, the contribution of SOI to the electronic state at a finite $T$ is examined.
\tk{When only SOI is considered, i.e., $\lambda_U = 0$, a metallic band appears owing to the edge state, as shown in Appendix A. 
In this case, the insulating behavior at low $T$ of $\alpha$-(BETS)$_2$I$_3$ cannot be explained.
In the following, we investigate the effects of SOI in the presence of \tk{on-site} Coulomb interactions \tk{$U_\alpha$}.
For simplicity, we set} $\lambda_{\rm SOI}\neq 0$ and $\lambda_U = 0.344$ as in the previous subsection.

\begin{figure}
\begin{centering}
\includegraphics[width=80mm]{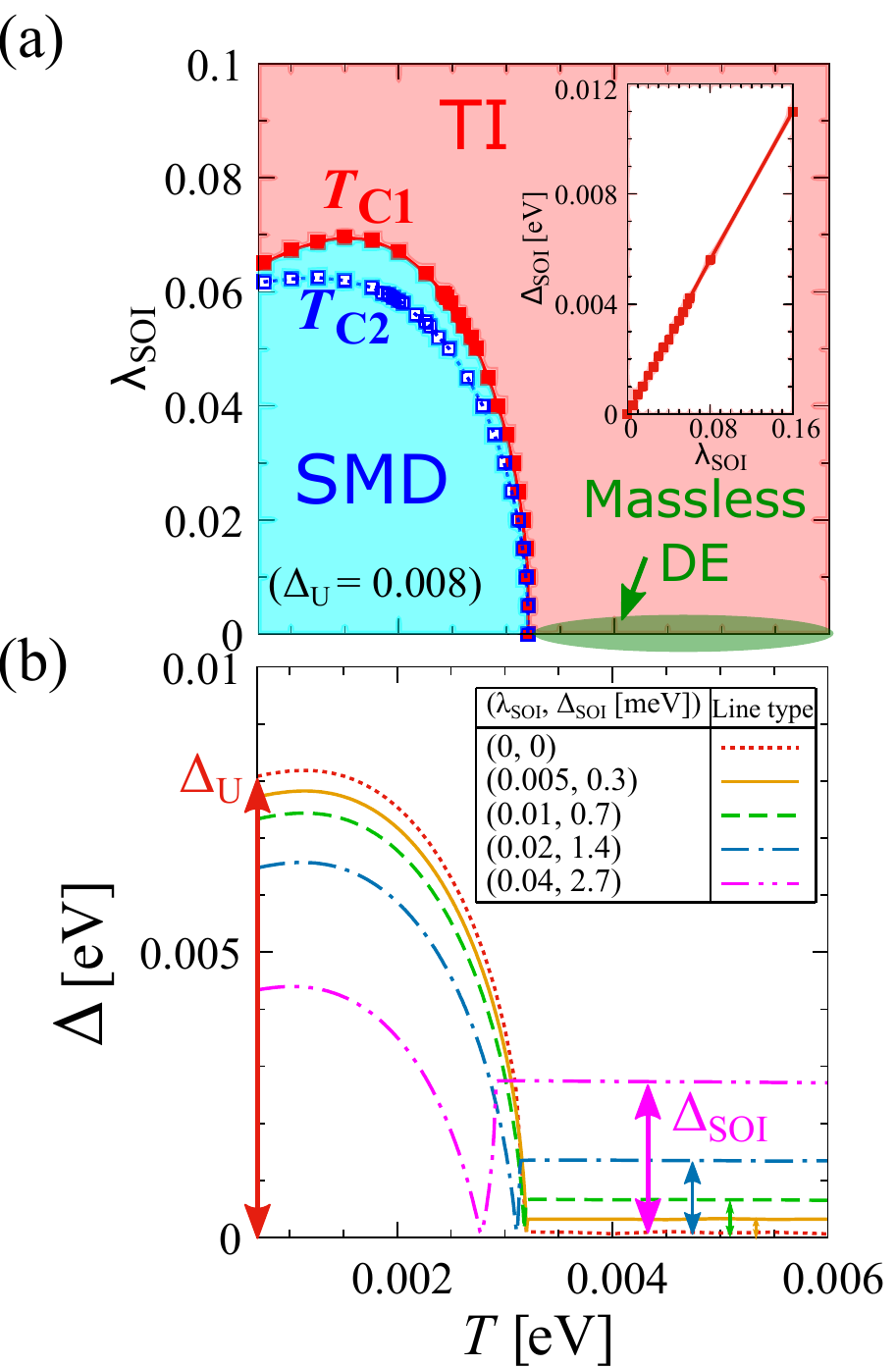}
\caption{\label{illustration}(Color online) (a) $\lambda_{\rm SOI}$-$T$ phase diagram.
\tk{SMD and a topological insulator (TI) indicate the SMD and topological insulator phases, respectively. 
\tk{Below $T_{\rm C2}$}, the spin Chern becomes zero, so \tk{the energy gap} $\Delta$ closes once. 
The blue dashed line shows the points at $\Delta=0$ in the SMD phase.}
\tk{The inset of (a) shows the $\lambda_{\rm SOI}$-dependence of the SOI gap $\Delta_{\rm SOI}$ at $\Delta_U = 0.008$.}
(b) $T$-dependence of the energy gap $\Delta$ at several values of \Erase{$\lambda_{\rm SOI}$ ($\lambda_U = 0.344$ fixed)}\tk{$\Delta_{\rm SOI}$ at $\Delta_U = 0.008$}.
}\label{Fig:SOIPDG}
\end{centering}
\end{figure}
Figure \ref{Fig:SOIPDG}(a) and (b) \tk{show} the $T-\lambda_{\rm SOI}$ phase diagram and the $T$-dependence of the energy gap $\Delta$ at several $\lambda_{\rm SOI}$ values, respectively.
Note that the value of the transfer integrals \Erase{is}\editage{has} the order of $10^{-1}$ eV (see Fig. \ref{Fig:2Dnetwork-TransferTable}), \Erase{so}\editage{therefore,} the magnitude of the SOI for $\lambda_{\rm SOI}=0.01$ is \Erase{about}\editage{approximately} 1 meV.
\tk{
Hereinafter, for convenience, we introduce two energy scales: SMD gap $\Delta_U$ and SOI gap $\Delta_{\rm SOI}$.
$\Delta_U = 0.008$ is defined as the value of the energy gap $\Delta$ at $(\lambda_U, \lambda_{\rm SOI}) = (0.344, 0)$ for $T = 0.0005$ (red solid arrow in Fig. \ref{Fig:ChargeSpinGap}(c) and Fig. \ref{Fig:SOIPDG}(b)).
$\Delta_{\rm SOI}$ is the value of $\Delta$ in $T > T_{\rm C1}$ which is associated with the energy scale of the SOI (magenta to orange solid arrows in Fig. \ref{Fig:SOIPDG}(b)).
The inset of Fig. \ref{Fig:SOIPDG}(a) shows the $\lambda_{\rm SOI}$-dependence of the SOI gap $\Delta_{\rm SOI}$ at $\Delta_U = 0.008$.
}
When $\lambda_{\rm SOI} > 0$ and $T > T_{\rm C1}$, the value of \Erase{$\Delta$}\tk{$\Delta_{\rm SOI}$} is finite, and the system becomes a topological insulator (TI) as described below.
It should be noted that\Erase{, in the} \editage{for }\Erase{large $\lambda_{\rm SOI}$ ($\lambda_{\rm SOI} > 0.015$)}\tk{$\Delta_{\rm SOI} > 0$ ($\lambda_{\rm SOI} > 0$)}, $\Delta$ exhibits a $V$-shaped $T$-dependence at $T < T_{\rm C1}$\tk{; i.e., }
$\Delta$ decreases to zero in $T_{\rm C2} < T < T_{\rm C1}$, becomes zero at $T = T_{\rm C2}$, and is finite again in $T < T_{\rm C2}$.
\tk{
As $\lambda_{\rm SOI}$ ($\Delta_{\rm SOI}$) is increased, $T_{\rm C1}$ gradually decreases and reaches to zero.
The SMD phase vanishes in $\lambda_{\rm SOI} > 0.07$ ($\Delta_{\rm SOI} \simeq 0.005$) and a quantum phase transition can occur when such a large SOI exists.
However, this value is more than twice the SOI value estimated by first-principles calculation \cite{TsumurayaSuzumura}.
}

%
\begin{figure*}[tb]
\begin{centering}
\includegraphics[width=150mm]{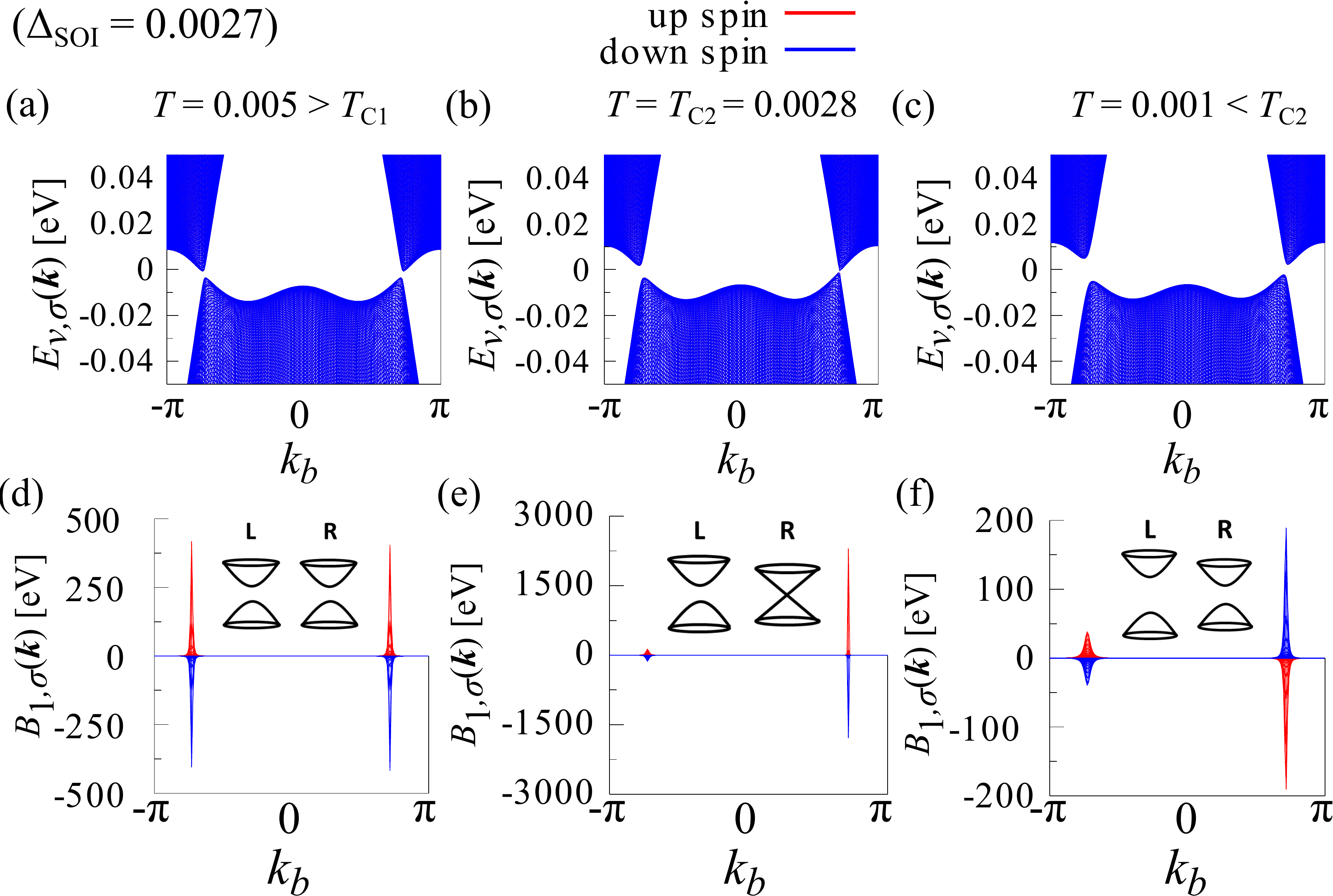}
\caption{\label{illustration}(Color online) Energy eigenvalues $E_{\nu,\sigma}({\bf k})$ near the Fermi energy and Berry curvature $B_{1,\sigma}({\bf k})$ at \Erase{$(\lambda_U, \lambda_{\rm SOI}) = (0.344, 0.04)$}\tk{$(\Delta_U, \Delta_{\rm SOI}) = (0.008, 0.0027)$} for the following three cases: (a) and (d): $T = 0.0050 > T_{\rm C1}=0.0032$, (b) and (e): $T = T_{\rm C2} = 0.0028$, and (c) and (f): $T = 0.0010 < T_{\rm C2}$. 
}\label{Fig:Band-BC-SOI}
\end{centering}
\end{figure*}
%
Figure \ref{Fig:Band-BC-SOI} shows the energy band $E_{\nu,\sigma}({\bf k})$ near the Fermi energy and Berry curvature $B_{1,\sigma}({\bf k})$ at \Erase{$(\lambda_U, \lambda_{\rm SOI}) = (0.344, 0.04)$}\tk{$(\Delta_U, \Delta_{\rm SOI}) = (0.008, 0.0027)$ ($(\lambda_U, \lambda_{\rm SOI}) = (0.344, 0.04)$)} in the following three cases: 
$T = 0.005 > T_{\rm C1}$\Erase{Comment: Please add the value of $T_{\rm C1}$.} [Figs. \ref{Fig:Band-BC-SOI}(a) and (d)], $T = T_{\rm C2} = 0.0028$ [Figs. \ref{Fig:Band-BC-SOI}(b) and (e)], and $T = 0.001 < T_{\rm C2}$ [Figs. \ref{Fig:Band-BC-SOI}(c) and (f)]. 
First\Erase{ly}, when $T = 0.005 > T_{\rm C1}$, the time-reversal symmetry exists, and the SOI gap opens at the Dirac point [Fig. \ref{Fig:Band-BC-SOI}(a)]. 
In this case, the sign of $B_{1,\sigma}({\bf k})$ is inverted according to the spin components, as illustrated in Fig. \ref{Fig:Band-BC-SOI}(d), and the system becomes the TI because the spin Chern number defined by $Ch_{S} \equiv Ch_{\uparrow} - Ch_\downarrow$ becomes $1$. 

Thereafter, in $T_{\rm C2} < T < T_{\rm C1}$ [Figs. \ref{Fig:Band-BC-SOI}(b) and (e)], the time-reversal symmetry is broken. 
\Erase{So}\editage{Hence,} $B_{1,\sigma}({\bf k})$ \tk{has} peaks \tk{with} different magnitudes according to the left and right valleys, and the spin Chern number \Erase{becomes}\editage{has} a real finite value. 
\tk{At $T = T_{\rm C2}$}, the sign of $B_{1,\sigma}({\bf k})$ in one valley is inverted corresponding to \tk{$\Delta=0$} at one valley. 
Finally, for $T < T_{\rm C2}$, gaps of different sizes are opened [Fig. \ref{Fig:Band-BC-SOI}(c)]. 
These behaviors in $T < T_{\rm C1}$ originate from the competition between the contributions of the spin order and SOI~\cite{Ezawa, Mong, Zheng, Miyakoshi, Cao, Hohenadler, Sekine, Jiang}. 
Moreover, as the sign of the $B_{1,\sigma}({\bf k})$ in one valley has \tk{been} already inverted at $T = T_{\rm C2}$, the spin Chern number is zero in this region [Figs. \ref{Fig:Band-BC-SOI}(c) and (f)].

\subsection{DC and optical conductivities}
In this subsection, \Erase{$\lambda_U$}\tk{$\Delta_U$} is fixed \tk{at} \Erase{$0.344$}\tk{0.008} as in the previous section, and the $T$ and SOI effects on the DC and optical conductivities are \tk{investigated}.

%
\begin{figure}
\begin{centering}
\includegraphics[width=75mm]{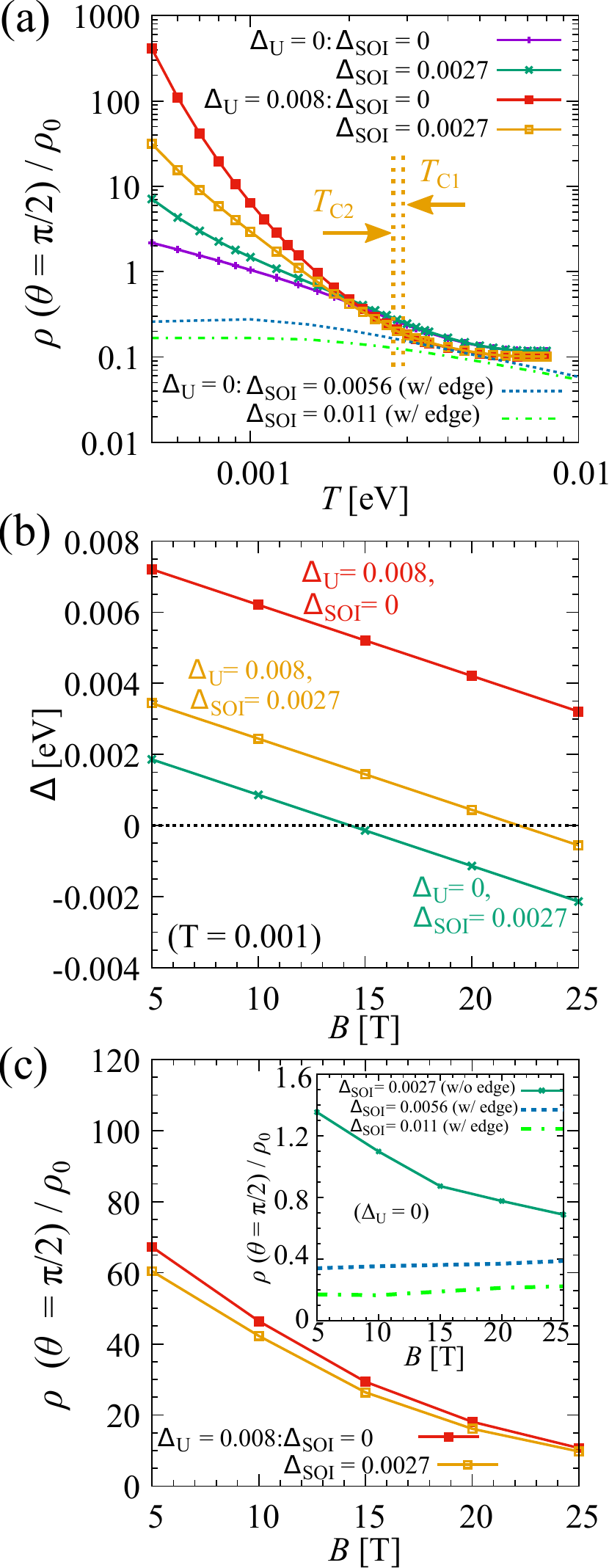}
\caption{\label{illustration} (Color online) 
(a) $T$-dependence of the $a$-axial DC resistivity $\rho(\theta=\pi/2)/\rho_0$ in units of the reciprocal of the universal conductivity $\rho_0\equiv 1/\sigma_0 = (4e^2/\pi h)^{-1}$ (solid lines).
The {results}\Erase{$b$-axial DC resistivity $\rho(\theta=0)/\rho_0$} at \Erase{$\lambda_U=0$}\tk{$(\Delta_U, \Delta_{\rm SOI}) = (0, 0.0056)$ (dashed line), and $(0, 0.011)$ (dotted chain line)}\Erase{for $\lambda_{\rm SOI}=0.08$ (dashed line) and $\lambda_{\rm SOI}=0.16$ (dotted chain line)} \tk{under the cylinder boundary condition} are also plotted.
Here, note that the slight increase in the resistivity at \Erase{$(\lambda_U, \lambda_{\rm SOI}) = (0, 0)$}\tk{$(\Delta_U, \Delta_{\rm SOI}) = (0, 0)$} near the lowest-$T$ is originated from the artificial gap by the accuracy limit of the numerical calculation.
(b) and (c) The in-plane magnetic-field $B$ dependence of (b) the energy gap $\Delta$ and (c) $\rho(\theta=\pi/2)/\rho_0$ at $T=0.001$ for several parameter sets of \Erase{($\lambda_U, \lambda_{\rm SOI}$)}\tk{$(\Delta_U, \Delta_{\rm SOI})$}.
The inset shows the $B$-dependence of {$\rho(\theta=\pi/2)/\rho_0$}\Erase{$\rho(\theta=0)/\rho_0$} for \tk{$(\Delta_U, \Delta_{\rm SOI}) = (0, 0.0056)$ (dashed line), and $(\Delta_U, \Delta_{\rm SOI}) = (0.008, 0.011)$ (dotted chain line)}.
}\label{Fig:Rho-Gap-vs-TandB}
\end{centering}
\end{figure}
%
The $T$-dependence of the $a$-axial DC resistivity $\rho(\theta=\pi/2)/\rho_0$ for \Erase{$\lambda_U=0$, $0.344$ and $\lambda_{\rm SOI}=0$, $0.04$}\tk{$(\Delta_U, \Delta_{\rm SOI}) = (0, 0)$, $(0, 0.0027)$, $(0.008, 0)$, and $(0.008, 0.0027)$} is plotted in Fig. \ref{Fig:Rho-Gap-vs-TandB}(a) as solid lines.
When only the SOI is considered \tk{($(\Delta_U,\Delta_{\rm SOI}) = (0, 0.0027)$)}, \Erase{as indicated by the solid lines at $\lambda_U=0$ and $\lambda_{\rm SOI}=0.04$, }the system becomes the TI, in which the SOI gap \tk{$\Delta_{\rm SOI}$} is opened at the Dirac point and $\rho(\theta=\pi/2)/\rho_0$ increases at quite low $T$ as $T$ \Erase{decreases}\editage{is decreased}.
Moreover, when considering the on-site Coulomb interaction, $\rho(\theta=\pi/2)/\rho_0$ increases below the phase transition temperature owing to the spin order gap.
However, as a result of the finite energy width owing to $-df/d\omega$ and the gentle function, such as $\sqrt{T}$ of the \Erase{spin order}\tk{energy} gap \tk{$\Delta$} [see Eq.(12) and Fig. \ref{Fig:ChargeSpinGap}(c)], $\rho(\theta=\pi/2)/\rho_0$ does not increase suddenly near the SMD phase transition temperature $T_{\rm C1}$.
When both the on-site Coulomb interaction $U$ and SOI are taken into account, the spin order gap is suppressed by the SOI.
\Erase{So}\editage{Thus,} $\rho(\theta=\pi/2)/\rho_0$ is suppressed at low $T$.

Here, note that in Fig. \ref{Fig:Rho-Gap-vs-TandB}(a), we also plot the $T$-dependence of $\rho(\theta=\pi/2)/\rho_0$ at \Erase{$\lambda_U=0$ for $\lambda_{\rm SOI}=0.08$ (dashed line) and $\lambda_{\rm SOI}=0.16$}\tk{$(\Delta_U, \Delta_{\rm SOI}) = (0, 0.0056)$ (dashed line), and $(0, 0.011)$} (dotted chain line) obtained by the calculation using the cylindrical boundary condition.
When only the SOI exists in the system with edge, the helical edge state appears, and {$\rho(\theta=\pi/2)/\rho_0$}\Erase{the DC resistivity} saturates, as shown by {these lines}\Erase{this line}.
\editage{Owing to the edge conduction,} the value of {$\rho(\theta=\pi/2)/\rho_0$}\Erase{the DC resistivity} \Erase{due to the edge conduction }has no \Erase{serious}\editage{significant} change even when we consider {large }SOI\Erase{(approximately 10 meV when $\lambda_{\rm SOI}=0.16$)}.
\Erase{So}\editage{Therefore,} we cannot explain the divergent increase of the DC resistivity observed in the experiment of $\alpha$-(BETS)$_2$I$_3$ when considering the SOI alone, and the edge state is robust (See Appendix A for details).

Figures \ref{Fig:Rho-Gap-vs-TandB}(b) and (c) represent the in-plane magnetic field $B$-dependence of the energy gap $\Delta$ and $\rho(\theta=\pi/2)/\rho_0$ for several values of \Erase{($\lambda_U$, $\lambda_{\rm SOI}$)}\tk{$(\Delta_U, \Delta_{\rm SOI})$}.
The energy band is split\Erase{ted} by $-{\rm sgn}(\sigma)\mu_{\rm B}B$ (see Eq. (1))\Erase{, so}\editage{.} 
\editage{Thus,} $\Delta(B)$ monotonically decreases as $B$ \Erase{increases}\editage{is increased} when calculating without edges.
As a result, in Fig. \ref{Fig:Rho-Gap-vs-TandB}(c) and the solid line in its inset,  {$\rho(\theta=\pi/2)/\rho_0$}\Erase{the DC resistivity} decreases as $B$ \Erase{increases}\editage{is increased}.
This result is consistent with the negative magnetoresistance observed in $\alpha$-(BETS)$_2$I$_3$ \cite{TajimaPriv}.
However, when considering the edge in the system, as shown by the dashed line and dotted chain line in the inset, $\rho(\theta=\pi/2)/\rho_0$ \Erase{shows}\editage{is} almost constant\Erase{ value}, owing to the edge conduction. 
\Erase{So}\editage{Hence,} we can not explain the negative magnetoresistance when considering the SOI alone.

%
\begin{figure}
\begin{centering}
\includegraphics[width=80mm]{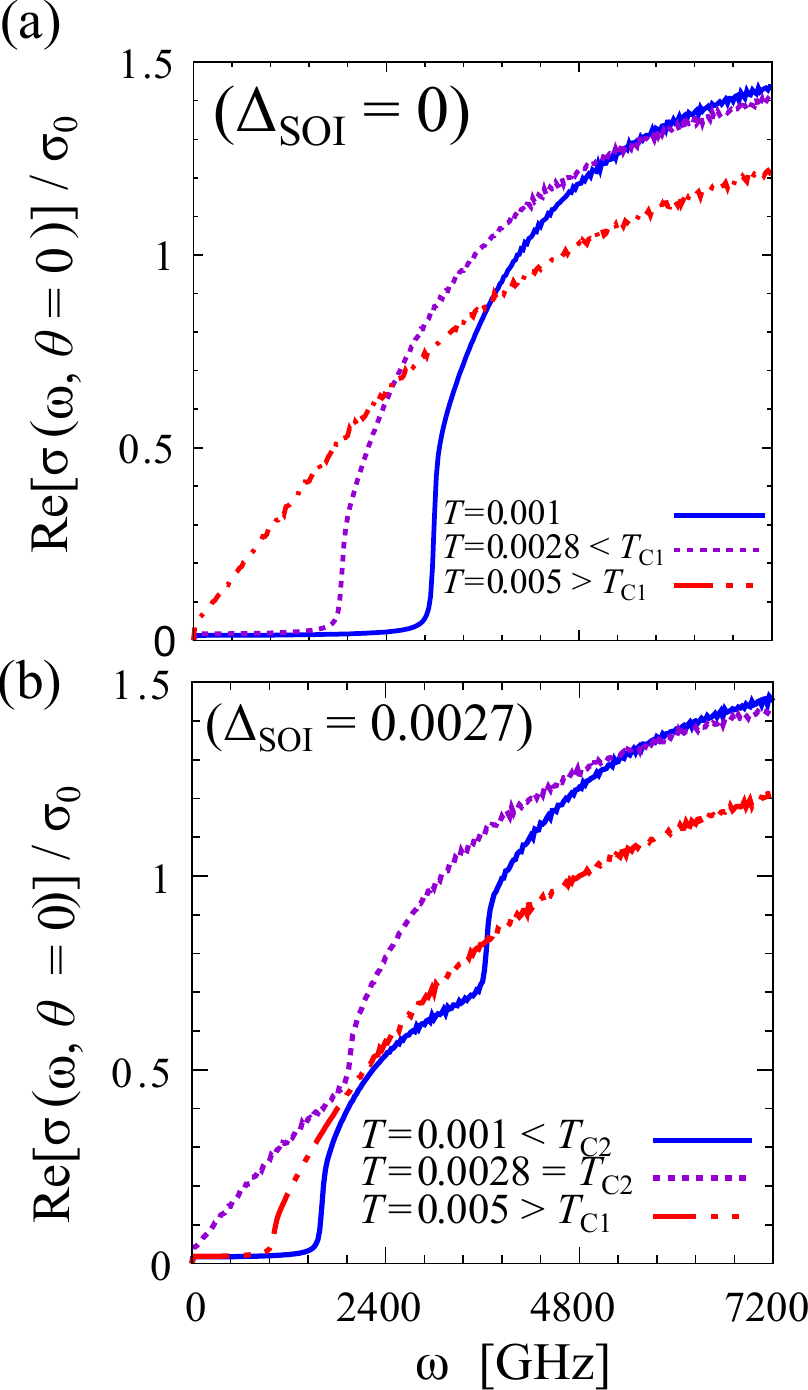}
\caption{\label{illustration} \Erase{Comment: the label in figure is wrong. Please change to Re$[\sigma(\omega,\theta=0)]/\sigma_0$.}(Color online) Real part of the optical conductivity along the $b$-axis ($\theta=0$) direction Re$[\sigma(\omega,\theta=0)]/\sigma_0$ in units of the universal conductivity $\sigma_0=4e^2/\pi h$ for (a) \Erase{$\lambda_{\rm SOI} = 0$}\tk{$\Delta_{\rm SOI} = 0$} and (b) \Erase{$\lambda_{\rm SOI} = 0.04$}\tk{$\Delta_{\rm SOI} = 0.0027$}.
$T$ is fixed \tk{at} $T = 0.005 > T_{\rm C1}=0.0032$ (dotted chain line), $T = 0.0028 < T_{\rm C1}$ (broken line), and $T = 0.001$ (solid line).
}\label{Fig:OPcnd-W}
\end{centering}
\end{figure}
%
Figures \ref{Fig:OPcnd-W}(a) and (b) \tk{show} the real part of the optical conductivity along the $b$-axis ($\theta=0$) direction Re$[\sigma(\omega,\theta=0)]/\sigma_0$ for $\lambda_{\rm SOI} = 0$ and \tk{$\Delta_{\rm SOI} = 0.0027$ ($\lambda_{\rm SOI} = 0.04$)} around $T = T_{\rm C1}$.
Re$[\sigma(\omega,\theta=0)]/\sigma_0$ \tk{shows clear differences} depending on the presence or absence of the SOI.
In $T = 0.005 > T_{\rm C1}$, Re$[\sigma(\omega,\theta=0)]/\sigma_0$ without the SOI \tk{has} a finite value at frequency $\omega = 0$, whereas that with the SOI remains zero until $\omega$ reaches approximately 960 GHz because of the finite SOI gap.
In $T < T_{\rm C1}$, Re$[\sigma(\omega,\theta=0)]/\sigma_0$ without the SOI becomes zero when the value of $\omega$ is smaller than the spin order gap $\Delta$, and increases abruptly in $\omega > \Delta$.
However, when the SOI \Erase{exists}\editage{is considered}, $\Delta$ exhibits a V-shaped $T$-dependence owing to the competition between the SMD and SOI, as indicated \tk{in} Fig. \ref{Fig:SOIPDG}(b).
As a result, Re$[\sigma(\omega,\theta=0)]/\sigma_0$ increases abruptly \Erase{twice}\editage{by two times} corresponding to the different $\Delta$s in the left and right valleys.
Furthermore, \tk{at} $T = T_{\rm C2}$, Re$[\sigma(\omega,\theta=0)]/\sigma_0$ with the SOI \tk{has} a finite value because \tk{$\Delta$} in the right valley is closed.

%
\begin{figure}
\begin{centering}
\includegraphics[width=\columnwidth]{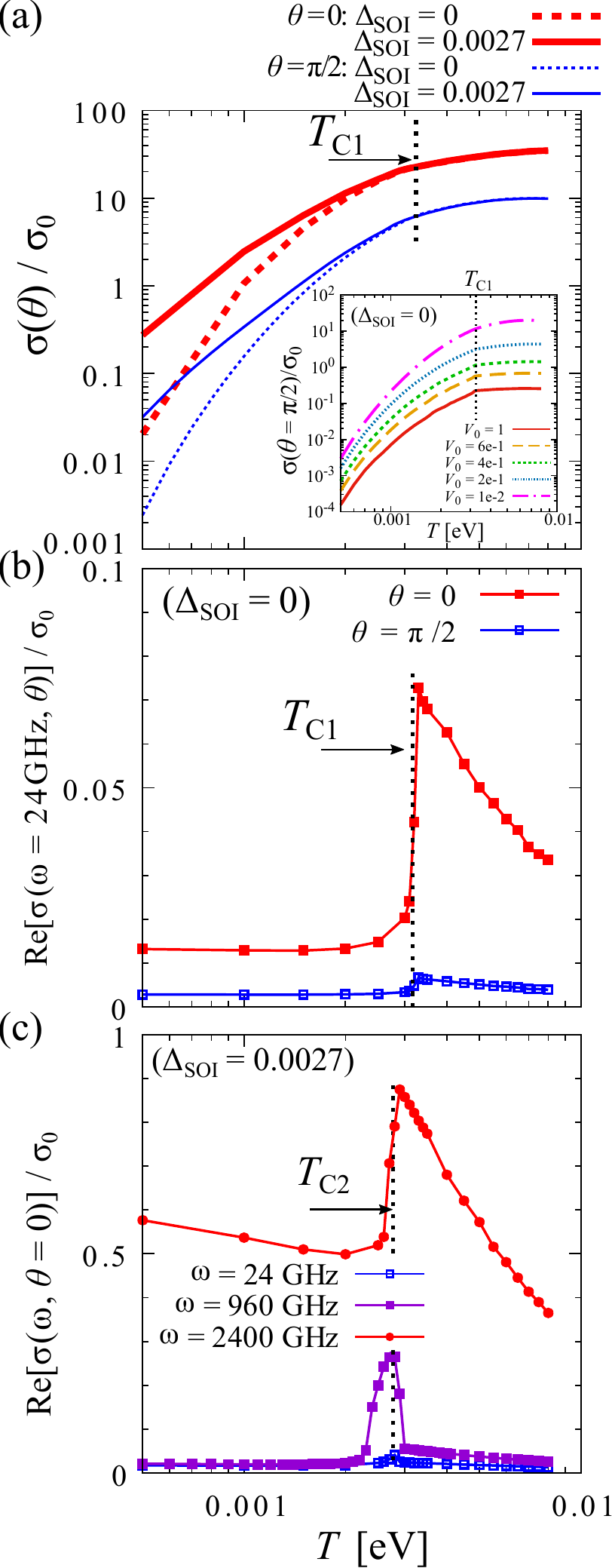}
\caption{\label{illustration}(Color online) $T$-dependence of (a) the DC conductivity $\sigma(\theta)/\sigma_0$, (b) real part of the optical conductivity Re$[\sigma(\omega=24{\rm GHz},\theta)]/\sigma_0$ at \Erase{$\lambda_{SOI} = 0$}\tk{$\Delta_{\rm SOI} = 0$}, and (c) real part of the optical conductivity Re$[\sigma(\omega,\theta=0)]/\sigma_0$ at \Erase{$\lambda_{\rm SOI} = 0.04$}\tk{$\Delta_{\rm SOI} = 0.0027$} in units of the universal conductivity $\sigma_0=4e^2/\pi h$ for fixed $\omega = 24, 960, 2400$ GHz.
\tk{Inset of (a) shows $T$-dependence of $\sigma(\theta)/\sigma_0$ at $\theta = \pi/2$, $\Delta_{\rm SOI} = 0$ for several values of the strength of impurity potential $V_0$.}
}\label{Cnd-T}
\end{centering}
\end{figure}
%
Figure \ref{Cnd-T}(a) \tk{shows} the $T$-dependence of the DC conductivity $\sigma(\theta)$ along the $b$-axis ($\theta=0$) and $a$-axis ($\theta=\pi/2$) directions.
$\sigma(\theta)$ decreases exponentially in $T < T_{\rm C1}$, but a clear discontinuous jump does not appear \tk{at $T= T_{\rm C1}$} because $\sigma(\theta)$ is influenced by the energy width of $-df/d\omega$, as indicated in Eq. (12).
\tk{
$T$-dependence of $\sigma(\theta)/\sigma_0$ at $\theta = \pi/2$, \tk{$\Delta_{\rm SOI} = 0$}\Erase{$\lambda_{\rm SOI} = 0$} for several values of the strength of impurity potential $V_0$ (see eq. (14)) are plotted in the inset of Fig. \ref{Cnd-T}(a).
The absolute value of $\sigma(\theta = \pi/2)/\sigma_0$ increases as $V_0$ is decreased from $V_0 = 1$ to $V_0 = 0.01 (=1e-2)$, but no discontinuous change appears at $T = T_C1$ for any $V_0$ value when the conductivity is calculated based on eqs. (12)-(15).
}
Figure \ref{Cnd-T}(b) shows the real part of the optical conductivity Re$[\sigma(\omega=24{\rm GHz},\theta)]$ in the absence of the SOI. 
As $T$ \Erase{decreases}\editage{is decreased}, in contrast to the DC conductivity, Re$[\sigma(\omega=24{\rm GHz},\theta)]$ increases gradually towards $T=T_{\rm C1}$ and decreases suddenly in $T < T_{\rm C1}$. 
The optical conductivity calculated by Eqs. (7) to (9) is considered as a direct transition in the inter-band at the same wavenumber and frequency $\omega$. 
Therefore, when $\Delta$ \Erase{appears}\tk{is finite} in $T < T_{\rm C1}$, the possible direct transition at the energy $\omega = 24$ GHz $\simeq 1$ eV disappears and Re$[\sigma(\omega=24{\rm GHz},\theta)]$ decreases sharply.
Finally, the $T$-dependence of Re$[\sigma(\omega,\theta)]$ in the presence of the SOI for several frequencies is plotted in Fig. \ref{Cnd-T}(c).
Re$[\sigma(\omega,\theta)]$ with the SOI \tk{has} a peak at $T=T_{\rm C2}$, where the gap of the right valley is closed.

\section{Summary and Discussion}

In this study, first\Erase{ly}, a Hubbard model was constructed as an effective model in the two-dimensional conduction plane of $\alpha$-(BETS)$_2$I$_3$ based on the synchrotron X-ray diffraction data at 30K under ambient pressure. 
We investigated the effects of the on-site Coulomb interaction $U$ and \Erase{the spin-orbit interaction (SOI)}\editage{SOI} at a finite temperature $T$ within the Hartree and $T$-matrix approximations to \Erase{elucidate}\editage{clarify} the insulating behavior observed in $\alpha$-(BETS)$_2$I$_3$ in the low $T$ \tk{region}. 

We found the phase transition between the weak \Erase{topological insulator (TI)}\editage{TI} phase and \Erase{the spin-ordered massive Dirac electron (SMD)}\editage{SMD} phase\Erase{ transition}.
In the SMD phase, the time-reversal symmetry is broken, but the spatial inversion and translational symmetries are conserved.
The SMD phase is not a conventional spin-ordered state, but exhibits the physical properties that reflect the wave functions of Dirac electrons. 
It is expected that the spin-valley Hall effect occurs\Erase{, since} \editage{because} the sign of the Berry curvature is reversed depending on the freedoms of the spin and valley. 
The SMD has the energy gap at the Dirac points, \Erase{while}\editage{whereas} the energy band in the bulk does not split in the spin degrees of freedom. 
The energy gaps of different sizes open in the left and right valleys \Erase{due}\editage{owing} to the competition between the SMD and SOI, as shown in the honeycomb lattice system in previous studies \cite{Ezawa, Mong, Zheng, Miyakoshi, Cao, Hohenadler, Sekine, Jiang}. 
Next, we calculated the $T$- and $B$-dependence\editage{s} of the \Erase{direct current (DC)}\editage{DC} resistivity. 
When considering the SOI alone and the system has edges, the helical edge state appears in the energy gap, and the DC resistivity saturates toward low $T$. 
The negative magnetoresistance does not appear in this case.
On the other hand, in the SMD phase, the DC resistivity increases divergently as $T$ \Erase{decreases}\editage{is decreased}, and there is no noticeable change near the SMD phase transition temperature $T_{\rm C1}$.
The DC resistivity exhibits the negative magnetoresistance, owing to the Zeeman split of the energy band.
Finally, it was shown that the $T$-dependence of the microwave (about $10^{-4}$ eV) conductivity shows clear changes at the vicinity of $T=T_{\rm C1}$. 

In recent magnetoresistivity measurements, a positive magnetoresistance and a negative magnetoresistance were observed at $T > 50$ K under in-plane and perpendicular magnetic fields, respectively.
This is the characteristic of the two-dimensional Dirac electron system \cite{TajimaPriv}.
On the other hand, a negative magnetoresistance appeared at $T < 50$ K under both in-plane and perpendicular magnetic fields \cite{TajimaPriv}. 
Furthermore, it was also pointed out that at $T < 50$ K, the Seebeck coefficient exhibits a nonmonotonic $T$-dependence \cite{TajimaPriv}. 
Those experimental results indicate that the electronic states change around 50 K. 
The TI-SMD transition shown in the present paper is consistent with the electric transport properties and the structure analysis observed in $\alpha$-(BETS)$_2$I$_3$ \cite{Inokuchi,KitouSawaTsumuraya,TajimaPriv}.
The existence of the TI-SMD transition can be directly confirmed by the microwave conductivity. 

\tk{
In the present study, the control parameter of SOI $\lambda_{\rm SOI}$ and spin $S_z$ are treated as constants for simplicity.
When we only discuss the qualitative behavior, the results shown in the present study (e.g. difference of the size in energy gaps between two Dirac cones in $T_{\rm C1} > T > T_{\rm C2}$, and TI phase in $T > T_{\rm C1}$) can be explained well in the range of this approximation and is robust for any other treatment of SOI because these behaviors are originated from the effects of on-site $U$ \cite{Ezawa, Mong, Zheng, Miyakoshi, Cao, Hohenadler, Sekine, Jiang}.
Time reversal symmetry (TRS) is conserved in TI phase, but antiferromagnetism induced by $U$ breaks the TRS \cite{Cao}, and causes the different size of energy gap between the left and right cones. Therefore, main result in this study is due to the effect of $U$, regardless of the detailed handling about SOI, so the approximation used in this study is sufficient to show the main results in our study.
However, more exactly, it is necessary to treat $\lambda_{\rm SOI}$ and spin $S_z$ as vector quantities in consideration of the anisotropy of SOI \cite{Winter} to have a quantitative discussion.
}

\tk{
When the spin order such as SMD phase appears, a clear change is expected to appear in the spin susceptibility.
In the NMR experiment for $\alpha$-(BETS)$_2$I$_3$ \cite{Hiraki}, no signs of magnetic transition have been observed near the insulating phase.
However, clear changes in the physical quantities in NMR (Knight shift and $1/T_1T$) originated from the SMD phase transition may be canceled out due to the nature of phase of the wave function in the Dirac electron systems.
This behavior can also be shown in Dirac electron systems such as an anisotropic square lattice model \cite{Katayama2006}.
}
The detailed analysis of the SMD phase and physical quantities of NMR \tk{in $\alpha$-(BETS)$_2$I$_3$} are \tk{currently in progress and}\Erase{to} \tk{will} be reported in \Erase{the other}\editage{another} paper.
The nonmonotonic $T$-dependence on the Seebeck coefficient of $\alpha$-(BETS)$_2$I$_3$ is also to be investigated in the future. 
When the time-reversal symmetry is broken by the SMD phase, the helical edge state due to the SOI is not protected, and the energy gap can open \cite{Biswas, Fang, Tokura}. 
Transport properties in the presence of impurities on the edges are to be investigated in the SMD phase with the SOI.

\tk{
The SMD phase is expected to be affected by the long-range Coulomb interaction and the spin fluctuation enhanced near the SMD phase transition.
To treat these effects, the calculation using the extended Hubbard model and the vertex correction\cite{YoshimiMaebashi, Raghu} should be performed and remains as future problems.
}

\begin{acknowledgments}
The authors would like to thank H. Sawa, T. Tsumuraya, and S. Kitou for valuable discussions and advices about the numerical calculation. 
We would also like to thank S. Onari, Y. Yamakawa, and H. Kontani for fruitful discussions. 
The computation in this work was conducted using the facilities of the Supercomputer Center, Institute for Solid State Physics, University of Tokyo. 
This work was supported by MEXT/JSPJ KAKENHI under grant numbers 19J20677, 19H01846, and 15K05166. 
\end{acknowledgments}

\appendix
\renewcommand{\thefigure}{\Alph{section}.\arabic{figure}}
\setcounter{figure}{0}
\section{Electrical resistivity if only spin-orbit interaction is considered\Erase{when considering SOI in the system with edges}}

%
\begin{figure}[t]
\begin{centering}
\includegraphics[width=0.8\columnwidth]{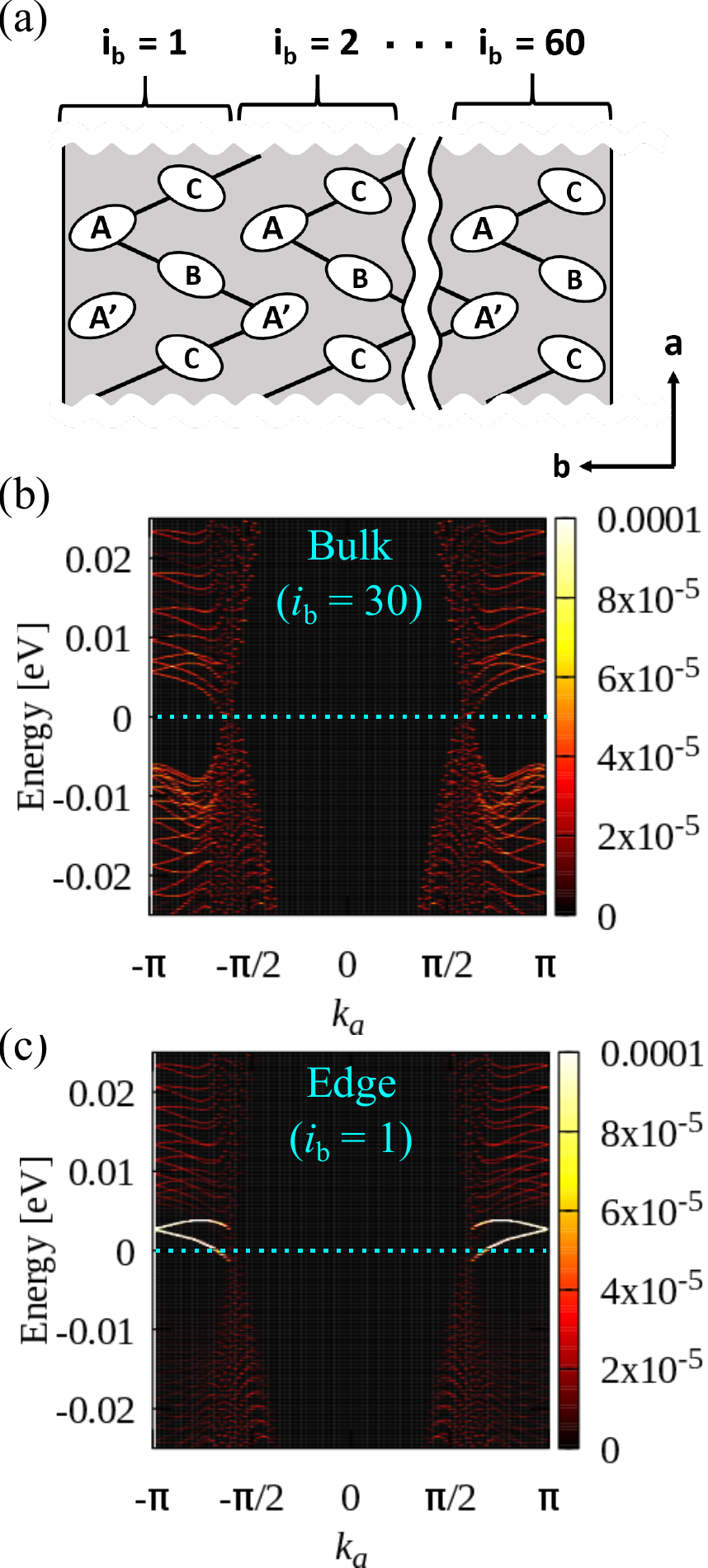}
\caption{\label{illustration}(Color online) (a) Schematic diagram of the cylindrical boundary condition imposed in the calculation.
The {left (right)}\Erase{bottom (top)} edge is formed by {sites A and A$'$ (B and C)}\Erase{sites A$'$ and B (A and C)}.
(b) and (c) Color plot of the spectral-weight {$\rho^S(i_b,k_a,\omega)$}\Erase{$\rho^S(i_a,k_b,\omega)$} near the Fermi energy set as the energy origin plotted for (b) {$i_b=30$}\Erase{$i_a = 15$} (bulk) and (c) {$i_b=1$ (left edge)}\Erase{$i_a = 30$ (top edge)}.
}\label{Fig:SpectralWeight}
\end{centering}
\end{figure}
%
In this appendix, we \tk{show} the results of \editage{the analysis of} the DC resistivity of $\alpha$-(BETS)$_2$I$_3$ when only the SOI is considered\Erase{considering SOI in the system with edges}.
To investigate the effects of the edge state on the DC resistivity, we impose the cylindrical boundary condition on the system, as illustrated in Fig. \ref{Fig:SpectralWeight}(a), and consider\Erase{ the on-site Coulomb interaction $U$ and} the term of SOI introduced in the main text and Ref \cite{Osada}.
The Fourier inverse transform is performed in the $a$\Erase{$b$}-axial direction and represented by the wavenumber {$k_a$}\Erase{$k_b$}, whereas the real space structure in the {$b$}\Erase{$a$}-axial direction is labeled by the coordinates of the unit cell {$i_b$}\Erase{$i_a$}.
The system size along the {$b$}\Erase{$a$}-axis is set to {$N_b = 60$}\Erase{$N_a = 30$}, as illustrated in Fig. \ref{Fig:SpectralWeight}(a), and thus, the Hamiltonian becomes a {$4N_b\times4N_b$}\Erase{$4N_a\times4N_a$} Hermitian matrix about each spin, which includes the information of the sublattice $\alpha$ ($=$A, A$'$, B, and C) and the unit cell coordinate {$i_b$ ($=1,\cdots N_b=60$)}\Erase{$i_a$ ($=1,\cdots N_a=30$)}.

As a result of the numerical diagonalization, we obtain {240}\Erase{120} energy eigenvalues {$E_{\nu,\sigma}(k_a)$ ($E_{1,\sigma}(k_a) < E_{2,\sigma}(k_a) < \cdots < E_{240,\sigma}(k_a)$)}\Erase{$E_{\nu,\sigma}(k_b)$ ($E_{1,\sigma}(k_b) < E_{2,\sigma}(k_b) < \cdots < E_{120,\sigma}(k_b)$)} and the unitary matrix {$d_{i_b,\alpha,\nu,\sigma}(k_a)$}\Erase{$d_{i_a,\alpha,\nu,\sigma}(k_b)$}.
Here, we introduce the spectral weight in each unit cell defined as
{
\begin{eqnarray}
\rho^S(i_b,k_a,\omega)&=&\sum_{\nu,\sigma}\left|d_{\i_b,\alpha,\nu,\sigma}(k_a)\right|^2\nonumber\\
&&\times\delta(\hbar\omega-E_{\nu,\sigma}(k_a)).
\end{eqnarray}
}
\Erase{
\begin{eqnarray}
\rho^S(i_a,k_b,\omega)&=&\sum_{\nu,\sigma}\left|d_{\i_a,\alpha,\nu,\sigma}(k_b)\right|^2\nonumber\\
&&\times\delta(\hbar\omega-E_{\nu,\sigma}(k_b)).
\end{eqnarray}
}
Figures \ref{Fig:SpectralWeight}(b) and (c) describe the {$\rho^S(i_b,k_a,\omega)$}\Erase{$\rho^S(i_a,k_b,\omega)$} for {$i_b=30$}\Erase{$i_a=15$} (bulk) and {$i_b=1$ (left edge)}\Erase{$i_a=30$ (top edge)} for the parameters of \Erase{$(T, \lambda_U, \lambda_{\rm SOI}) = (0, 0, 0.08)$}\tk{$(T, \Delta_U, \Delta_{\rm SOI}) = (0, 0, 0.0056)$ (when considering the SOI alone)}.
Although {$\rho^S(30,k_a,\omega)$}\Erase{$\rho^S(i_a=15,k_b,\omega)$} in Fig. \ref{Fig:SpectralWeight}(b) is spread weakly over the whole energy range, {$\rho^S(1,k_a,\omega)$}\Erase{$\rho^S(i_a=30,k_b,\omega)$} \Erase{has}\editage{is} quite large\Erase{ value} near the Fermi energy, as shown in Fig. \ref{Fig:SpectralWeight}(c), owing to the existence of a helical edge state protected by the time-reversal symmetry in the system.
Therefore, the conduction channel of this edge state becomes dominant at $T=0$.

%
\begin{figure}[tb]
\begin{centering}
\includegraphics[width=0.8\columnwidth]{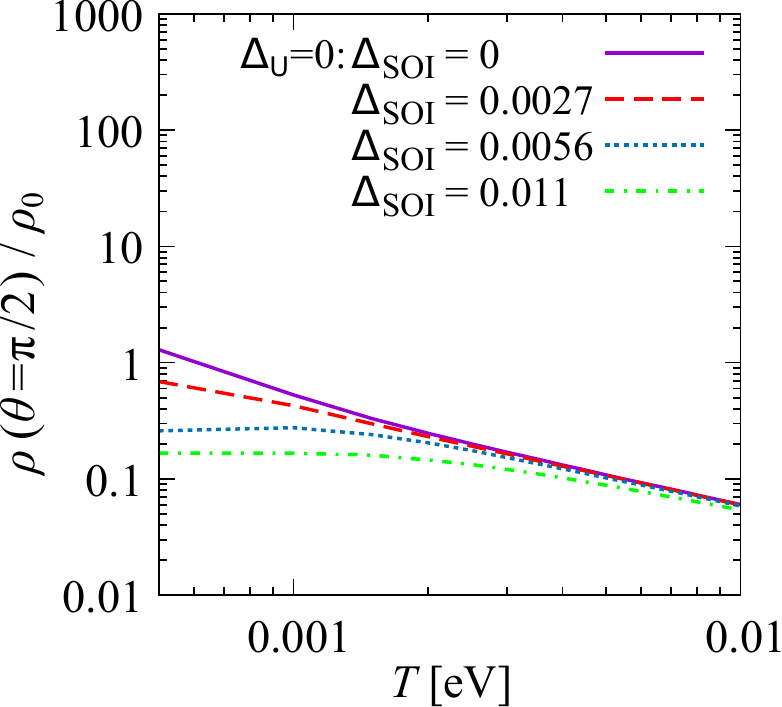}
\caption{\label{illustration}(Color online) $T$-dependence of the DC resistivity along the {$a$-axis ($\theta=\pi/2$)}\Erase{$b$-axis ($\theta=0$)} in units of the reciprocal of the universal conductivity \Erase{$\sigma_0$}\tk{$\rho_0\equiv 1/\sigma_0$} at \Erase{$\lambda_U=0$ for $\lambda_{\rm SOI} = 0$, 0.04, 0.08, and 0.16}\tk{$(\Delta_U, \Delta_{\rm SOI}) = (0, 0)$, $(0, 0.0027)$, $(0, 0.0056)$, and $(0, 0.011)$}.
}\label{Fig:Rho-T-Edge}
\end{centering}
\end{figure}
%
Figure \ref{Fig:Rho-T-Edge} shows the $T$-dependence of the DC resistivity for \Erase{$\lambda_{\rm SOI} = 0$, 0.04, 0.08, and 0.16}\tk{$(\Delta_U, \Delta_{\rm SOI}) = (0, 0)$, $(0, 0.0027)$, $(0, 0.0056)$, and $(0, 0.011)$}.
When the SOI is considered in the bulk, as calculated in the main text, the energy gap opens at the Dirac point, and the system becomes an insulator. 
However, when considering the SOI in a system with edges, the helical edge state appears in the vicinity of the Fermi energy owing to the band crossing between the up and down spin bands, so that it does not actually become an insulator.
Note that the slight increase in the resistivity at \Erase{$\lambda_{\rm SOI}=0$}\tk{$\Delta_{\rm SOI}=0$} near the lowest $T$ results from the energy gap associated with the finite-size effect.
\tk{
The edge state caused by SOI is topologically protected.
On the other hand, edge state which is not protected and depends on the edge setting also appears in some cases.
For instance, in the $\alpha$-type organic conductors, it is suggested that when the edge setting is symmetric, the edge state appears in the gapless band\cite{Omori2017, Ohki}.
Therefore, when such an edge state exists and the spin order by $U$ occurs, it is expected that the edge state associated with the AF at the edge appears in the gapless band and edge conduction occurs.
}





\end{document}